%% file: ratio07.tex
\documentclass[12pt]{article}

\usepackage{changebar}
\usepackage{url}
\usepackage{amsmath}
\usepackage{amssymb}
\usepackage{handbookbib}

\usepackage{color}
\usepackage{graphics}

\newcommand{\GR}{GR}
\newcommand{\LR}{LR}
\newcommand{\GS}{GS}
\newcommand{\LS}{LS}
\newcommand{\MGS}{MGS}
\newcommand{\MLS}{MLS}

\include{ecmds}

\def\smallromani{\renewcommand{\theenumi}{\roman{enumi}}
\renewcommand{\labelenumi}{(\theenumi)}}

\newcommand{\Proof}{\NI
                    {\bf Proof.}\ }
\newtheorem{theorem}{Theorem}
\newtheorem{defined}{Definition}
\newenvironment{definition}{\begin{defined} \rm}{\end{defined}}
\newtheorem{exa}{Example}
\newenvironment{example}{\begin{exa} \rm}{\end{exa}}

\newtheorem{lemma}{Lemma}
\newtheorem{corollary}{Corollary}

\newtheorem{note}{Note}

\usepackage{amsfonts}
\usepackage{latexsym}
\usepackage{alltt}

\title{The Many Faces of Rationalizability}

\author{Krzysztof R. Apt \\
\emph{CWI, Amsterdam, the Netherlands} \\
and \emph{University of Amsterdam}
}

\begin{document}

\date{}

\maketitle

\begin{abstract}
  The rationalizability concept was introduced in \cite{Ber84} and
  \cite{Pea84} to assess what can be inferred by rational players in a
  non-cooperative game in the presence of common knowledge.  However,
  this notion can be defined in a number of ways that
  differ in seemingly unimportant minor details.  We shed light on
  these differences, explain their impact, and clarify for which games
  these definitions coincide.  
  
  Then we apply the same analysis to explain the differences and
  similarities between various ways the iterated elimination of
  strictly dominated strategies was defined in the literature.  This
  allows us to clarify the results of \cite{DS02} and \cite{CLL05} and
  improve upon them. We also consider the extension of these results
  to strict dominance by a mixed strategy.

  Our approach is based on a general study of the operators on
  complete lattices.  We allow transfinite iterations of the
  considered operators and clarify the need for them.  The advantage
  of such a general approach is that a number of results, including
  order independence for some of the notions of rationalizability and
  strict dominance, come for free.  
\end{abstract}




\section{Introduction}
\label{sec:introduction}

\subsection{Motivation}
\label{subsec:motivation}

Rationalizability was introduced in \cite{Ber84} and \cite{Pea84} to
formalize the intuition that players in non-cooperatives games act by
having common knowledge of each others' rational behaviour.
Rationalizable strategies are then defined as a limit
of an iterative process in which one repeatedly removes the strategies
that are never best responses to the beliefs held about the other
players. 

To better understand the rationale for the research here reported consider the
following example.

\begin{example} \label{exa:Bertrand1}\emph{Bertrand competition}.

Consider a version of Bertrand competition between two firms
in which the marginal costs are 0 and in which
the range of possible prices is the left-open real interval $(0, 100]$.
So in this game $H$ there are two players, each with the set $(0, 100]$ of strategies.
We assume that the demand equals $100 - p$, where $p$ is the lower price and
that the profits are split in case of a tie.
So the payoff functions are defined by:

\[
\begin{array}{l}
p_1(s_1, s_2) := \left\{ 
\begin{tabular}{ll}
$s_1 (100 - s_1) $ &  \mbox{if $s_1 < s_2$} \\[2mm]
$\dfrac{s_1 (100 - s_1)}{2} $ &  \mbox{if $s_1 = s_2$} \\[2mm]
0 &  \mbox{if $s_1 > s_2$} 
\end{tabular}
\right .  \\
\\
p_2(s_1, s_2) := \left\{ 
\begin{tabular}{ll}
$s_2 (100 - s_2) $ &  \mbox{if $s_2 < s_1$} \\[2mm]
$\dfrac{s_2 (100 - s_2)}{2} $ &  \mbox{if $s_1 = s_2$} \\[2mm]
0 &  \mbox{if $s_2 > s_1$} 
\end{tabular}
\right . 
\end{array}
\]

This game has no Nash equilibrium (in pure strategies).

Consider now each player's best responses to the
strategies of the opponent.
Since $s_1 = 50$ maximizes the value of $s_1 (100 - s_1)$ in the
interval $(0, 100]$, the strategy 50 is the unique best response 
of the first player to
any strategy $s_2 > 50$ of the second player.  Further, no strategy is
a best response to a strategy $s_2 \leq 50$.  By symmetry the same
holds for the strategies of the second player.

This eliminates for each player each strategy different than 50 and reduces the original
game to the game $G := (\C{50}, \C{50}, p_1, p_2)$ 
in which each player has just one strategy, 50.

There are now two natural ways to proceed. If we adopt the approach of
\cite{Pea84}, we should now focus on the \emph{current} game $G$ and note that
$s_1 = 50$ is a best response in $G$ to $s_2 = 50$ and symmetrically
for the second player. So the iterated elimination of never best
responses stops and the outcome is $G$.

However, if we adopt the approach of \cite{Ber84}, we should continue
to consider the best responses in the \emph{original} game $H$. Now in
$H$ the strategy $s_1 = 49$ is a better response to $s_2 = 50$ than
$s_1 = 50$ and symmetrically for the second player.  So in the second
round of elimination of never best responses both strategies 50 are
eliminated and we reach the empty game.

So depending on the procedure we adopt we obtain two different outcomes.

\HB
\end{example}

We shall return to this example in Section \ref{sec:rat}.  We shall
show there that using the first elimination procedure we can still
reach the empty game, if in each round \emph{only some} strategies are
removed.  This might suggest that both approaches are equivalent if we
do not insist on removing \emph{all} strategies in each round
(analogously to the case of iterated elimination of strictly dominated
strategies).  However, as we shall see, this statement is false.

So we see that the iterated elimination of best responses process can
be defined in two different ways. In fact, two other definitions can
be naturally envisaged.  Each of these four definitions captures the
original intuition in a meaningful way.  Indeed, all four operators
upon which these definitions rely yield the \emph{same} outcome when
applied to the original game.  The differences arise when these
operators are iterated.

Let us move now to the case of strict dominance. 
Consider the following example.

\begin{example} \label{exa:strict} \emph{Production with a discontinuity}.

Consider a game $H$ with two players, each, as in the previous example
with the set $(0, 100]$ of strategies.
The payoff functions are defined now by:

\[
\begin{array}{l}
p_1(s_1, s_2) := \left\{ 
\begin{tabular}{ll}
$f_1(s_1, s_2)$ &  \mbox{if $(s_1, s_2) \neq (100,100)$} \\[2mm]
$0 $ &  \mbox{otherwise}
\end{tabular}
\right .  \\
\\
p_2(s_1, s_2) := \left\{ 
\begin{tabular}{ll}
$f_2(s_1, s_2)$ &  \mbox{if $(s_1, s_2) \neq (100,100)$} \\[2mm]
$0 $ &  \mbox{otherwise}
\end{tabular}
\right .  \\
\end{array}
\]
where each function $f_i$ is strictly increasing in the $i$th argument.
(A simple example is $f_i(s_1, s_2) := s_i$.)

A possible interpretation of this game is as follows.  The strategy of
a player represents the amount of his resource that
he chooses.  If each player 'overdoes it' and chooses the maximum
amount, the outcome is bad (0) for both of them. Otherwise each player
gets the outcome computed by his production function $f_i$.
Also this game has no Nash equilibrium.

Consider now each player's strategies that are not strictly dominated.
Clearly every strategy $s_i \neq 100$ is strictly dominated and 100 is
not strictly dominated. By eliminating all strictly dominated
strategies the original game reduces to the game $G := (\C{100},
\C{100}, p_1, p_2)$ in which each player has just one strategy, 100.
The payoff for each player to the joint strategy $(100, 100)$ is 0.

Again, there are
two natural ways to proceed. If we take the usual approach, adopted 
in numerous publications, we should focus on the \emph{current} game, $G$.
Since this game is solved (i.e., each player has just one strategy) the
iterated elimination of strictly dominated strategies, in short IESDS,
stops and the outcome is $G$.

However, if we adopt the approach of \cite[ pages 1264-1265]{MR90}, we should continue
and consider which strategies in the game $G$ are strictly dominated
(against the opponent strategies in $G$) in the \emph{original} game $H$. Now,
for each player $i$ each strategy 100 is strictly dominated by any
other strategy $s_i$ in the game $H$ as each $s_i$ yields a strictly
higher payoff against the strategy 100 of the opponent. So in the
second round of elimination of strictly dominated strategies both
strategies 100 are eliminated and we reach the empty game.

So also here, depending on the procedure we adopt we obtain two
different outcomes.  (A perceptive reader may notice that the
elimination of never best responses also reduces the original game to
$(\C{100}, \C{100}, p_1, p_2)$ and that in the original game for each
player each strategy $\neq 100$ is a better response to 100 than 100.
So also here the outcome of the iterated elimination of never best
responses depends on the adopted procedure.)
\HB
\end{example}

In fact, we show that also IESDS can be defined in four natural ways and
that the resulting outcomes differ.

\subsection{Approach and summary of results}

To analyze in a uniform way various ways of iterated
elimination of strategies here considered 
we consider arbitary operators on complete lattices and
their transfinite iterations.  This allows us to prove various results
concerning the rationalizability and IESDS by simply checking the
properties of underlying operators.  For example, order independence
for specific definitions turns out to be a direct consequence of the
fact that the underlying operator is monotonic.

Before we proceed let us clarify two, rather unusual aspects of our
approach.  First, the use of transfinite induction and ordinals in
reasoning about games is rare though not uncommon.  The following
illustrative examples come to our mind.  In \cite[ Chapter 7]{Bin07}
(and implicitly in the original version, \cite{Bin91}) a proof,
attributed to G. Owen, of the Minimax Theorem is given that is based
on transfinite induction.  Next, in \cite{Lip91} transfinite ordinals
are used in a study of limited rationality, while in \cite{Lip94} a
two-player game is constructed for which the $\omega$ (the first
infinite ordinal) and $\omega + 1$ iterations of the rationalizability
operator of \cite{Ber84} differ.  That is, $\omega$ iterations are
insufficient to reach a fixpoint. This motivates the author to study
transfinite iterations of this operator.  We shall return to this
matter in Section \ref{sec:st}.  In turn, \cite{HS98} show that in
general arbitrary ordinals are necessary in the epistemic analysis of
strategic games based on the partition spaces.  Finally, as argued in
\cite{CLL05}, the notion of IESDS \`{a} la \cite{MR90}, when used for
arbitrary games, also requires transfinite iterations of the
underlying operator.

The mathematical reason for the use of transfinite induction is that
the underlying operators, even if they are monotonic, are in general
not continuous.  So to reach a fixpoint, by the theorem of
\cite{Tar55} (which generalizes and strengthens the initial result of
\cite{Kna28}), one needs to consider iterations that continue beyond
$\omega$.

Second, we consider solution concepts based on the iterated
elimination of strategies that can yield the empty outcomes. This is
inherent in the nature of infinite games as it can then easily happen
that no strategy is a best response or that each strategy is strictly
dominated. The empty outcome only indicates
that in some games the players have no rational strategy to choose from.
Analogous problems arise if one adopts as a 
solution concept the set of Nash equilibria of a non-cooperative
game or the core of a cooperative game. In both cases these sets
can be empty. To quote from \cite{Aum85}:

\begin{quote}
My main thesis is that a solution concept should be judged more by
what it does than by what it is; more by its success in establishing
relationships and providing insights into the workings of the social
processes to which it is applied than by considerations of a \emph{a priori}
plausibility based on its definition alone.
\end{quote}

Our analysis allows us to conclude that in the case of arbitrary games
among four ways of defining rationalizability only two, the one due
\cite{Ber84} and its contracting version
(a notion explained in Section \ref{sec:reductions}), are meaningful. The crucial
feature of these two operators is that they refer to the best
responses in the initial game and \emph{not} the currently considered
game.  As a result these two operators are monotonic. However, under
the assumption
that to each belief in the initial game a best response exists
(assumption \textbf{B}), the iterations of all four operators coincide
even though the other two still are not monotonic.  We also explain
the need for transfinite iterations of the corresponding operators,
even for the games that satisfy assumption \textbf{B}.
Assumption \textbf{B} is in particular satisfied by the compact games with continuous
payoff functions, which explains why the reported differences were not
discussed in the literature.

We also apply the same analysis to the notion of strict dominance.  We
explain that the operator underlying the usual definiton is not
monotonic. This clarifies why this definition of strict dominance is
not order independent for arbitrary infinite games
(see \cite{DS02}).

In contrast, the version of the iterated elimination of strictly dominated
strategies used in \cite{MR90} and more recently in \cite[ Section
5.1]{Rit02} is monotonic.  The contracting version of it studied in
\cite{CLL05}. The authors of the latter paper show that the resulting
elimination procedure is stronger than the customary one, requires
transfinite iterations, does not remove any Nash equilibria, and is
order independent.  In our framework order independence of both
versions is a direct consequence of the monotonicity of the underlying
operators.

We also consider a natural condition \textbf{C}($\alpha$) of the
initial game, paramet\-rized by an ordinal $\alpha$, that formalizes
the statement that in each reduction reachable from the initial game
in $\alpha$ iterations every strictly dominated strategy has an
undominated dominator.  We show that $\fa \alpha$\textbf{C}($\alpha$)
implies order independence of the usual definition of strict
dominance.  Further, $\fa \alpha$\textbf{C}($\alpha$) implies that the
iterations of all four considered operators coincide.  Assumption $\fa
\alpha$\textbf{C}($\alpha$) is in particular satisfied by the finite
games.

Our formalization of the condition \textbf{C}($\alpha$) differs from
the one used in \cite{DS02} for which the authors showed that the
iterated elimination of strictly dominated strategies may fail to be
order independent in their sense.  (Our definitions of order
independence differ, since we allow transfinite iterations of the
underlying operators.)  We show that \textbf{C}($\omega$) implies
order independence, in the sense used in \cite{DS02}, of strict
dominance.  This yields a minor improvement of their result.
Finally, we explain how to extend these results to the case of
iterated elimination of strictly dominated strategies by a mixed
strategy.

\subsection{Plan of the paper}

The paper is organized as follows.  Standard concepts on strategic
games and examples of belief structures to which the results of this
paper apply are introduced in Section \ref{sec:games}.  Next, in
Section \ref{sec:reductions} some general results about monotonic and
contracting operators on complete lattices are established that provide a basis for our
approach.

In Section \ref{sec:st} we discuss two operators that underly the
definitions of rationalizability, including the one due to
\cite{Ber84}. Both are monotonic, so the revelant properties of these
operators and of their outcomes are direct consequences of the general
results established in Section \ref{sec:reductions}.

Then in Section \ref{sec:rat} we discuss two slightly different
operators defining a notion of rationalizability.  We conclude that
one of them cannot be meaningfully used to formalize the notion of
rationalizability and the other one, underlying the definition of
rationalizability due to \cite{Pea84}, leads to too weak conclusions
for a game modelling Bertrand competition for two firms.  On the
other hand, in Section \ref{sec:coincide} we show that the best
response property in the initial game implies that the iterations of
all four operators coincide.

Then in Sections \ref{sec:sd} and \ref{sec:sd1} we discuss in detail
four natural ways of defining iterated elimination of strictly
dominated strategies and analyze when the iterations of the
corresponding four operators coincide.  In Section \ref{sec:comp} we
compare our results with those of \cite{DS02} and clarify in what
sense we established a new order independence result.

Next, in Section \ref{sec:mixed} consider an extension of these
results to the case of strict dominance by a mixed strategy.  Finally,
in Section \ref{sec:conc} we assess the results of this paper
stressing that monotonicity and transfinite iterations are most
relevant for a study of reduction operators on games.

In the literature we found only one reference concerned with a similar
comparative analysis of the notion of rationalizability. \cite{Amb94}
studied the limited case of two-player games and beliefs being equal to
the strategies of the opponent, and showed that the finite
iterations of two operators concerned with the notion
of rationalizablity coincide for compact games with continuous
payoffs.  We clarify his result in Section \ref{sec:coincide}.

Parts of this research were reported in \cite{Apt05}.

\section{Game theoretic preliminaries}
\label{sec:games}
\subsection{Strategic games and their restrictions}

Let us move now to the subject of strategic games.
Given $n$ players ($n > 1$) we represent a
strategic game (in short, a game)
by a sequence
\[
(S_1, \LL, S_n, p_1, \LL, p_n),
\] 
where for each $i \in [1..n]$

\begin{itemize}
\item $S_i$ is the non-empty set of \oldbfe{strategies} 
available to player $i$,

\item $p_i$ is the \oldbfe{payoff function} for the  player $i$, so
$
p_i : S_1 \times \LL \times S_n \myra \cal{R},
$
where $\cal{R}$ is the set of real numbers.
\end{itemize}

Given a sequence of sets of strategies $S_1, \LL, S_n$ and
$s \in S_1 \times \LL \times S_n$ we denote the $i$th element of $s$ by $s_i$ and
use the following standard notation:

\begin{itemize}
\item $s_{-i} := (s_1, \LL, s_{i-1}, s_{i+1}, \LL, s_n)$,


\item $S_{-i} := S_1 \times \LL \times S_{i-1} \times S_{i+1} \times \LL \times S_n$.

\end{itemize}
We denote the strategies of player $i$ by $s_i$, possibly with some superscripts.

We say that $G := (S_1, \LL, S_n)$ is a
\oldbfe{restriction} of a game $H := (T_1, \LL, T_n,$ $p_1, \LL,p_n)$ if
each $S_i$ is a (possibly empty) subset of $T_i$.
We identify the restriction $(T_1, \LL, T_n)$ with $H$.
The restrictions are naturally ordered by the componentwise set inclusion:
\[
\mbox{$(S_1, \LL, S_n) \sse (S'_1, \LL, S'_n)$ iff $S_i \sse S'_i$ for all $i \in [1..n]$.}
\]

If some $S_i$ is empty, $(S_1, \LL, S_n, p_1, \LL,p_n$ 
(we identify here each $p_i$ with its restriction to the smaller domain)
is not a game and the references to
$p_j(s)$ (for any $j \in [1..n]$) are incorrect, so we shall need to be
careful about this.  If all $S_i$ are empty, we call $G$ an
\oldbfe{empty restriction}.

\subsection{Belief structures}
\label{subsec:beliefs}

Throughout the paper $H : = (T_1, \LL, T_n, p_1, \LL,p_n)$ is a
\emph{fixed} game.  

Our intention is to explain various concepts of rationalizability
abstracting from specific sets of beliefs that are assumed. Therefore
we only assume that each player $i$ in the game $H$ has some further
unspecified non-empty set of beliefs ${\cal B}_{i}$ about his
opponents and that each payoff function $p_i$ can be modified to an
\oldbfe{expected payoff} function $p_i : S_i \times {\cal B}_{i} \myra
\cal{R}$.

In what follows we also assume that each set of beliefs ${\cal B}_{i}$
of player $i$ in $H$ can be \emph{narrowed} to any restriction $G$ of
$H$. We denote the outcome of this \oldbfe{narrowing} of ${\cal
  B}_{i}$ to $G$ by ${\cal B}_{i} \stackrel{.}{\cap} G$.  The 
set ${\cal B}_{i} \stackrel{.}{\cap} G$ can be viewed as the set of
beliefs of player $i$ in the restriction $G$.  We call then the pair $({\cal B},
\stackrel{.}{\cap})$, where ${\cal B} := ({\cal B}_{1}, \LL, {\cal
  B}_{n})$, a \oldbfe{belief structure} in the game $H$.

The following natural property of a belief structure $({\cal B},
\stackrel{.}{\cap})$ in $H$ will be relevant.

\begin{description}
\item[A] If $G _1\sse G_2 \sse H$, then for all $i \in [1..n]$, 
${\cal B}_{i} \stackrel{.}{\cap} G_1 \sse {\cal B}_{i} \stackrel{.}{\cap} G_2$.
\end{description}

This property simply states that for each player the set of his
beliefs in a restriction $G_1$ of $G_2$ is a subset of the set of his
beliefs in $G_2$.

The following four belief structures were considered in the
literature.  Given a finite non-empty set $A$ we denote here by
$\Delta A$ the set of probability distributions over $A$ and by
$\Delta^{\! \circ} A$ the set of probability distributions over $A$ that
assign a positive probability to each element of $A$.

\begin{enumerate} \smallromani
\item ${\cal B}_{i} := T_{-i}$ for $i \in [1..n]$.

So beliefs are joint pure strategies of the opponents
(usually called \oldbfe{point beliefs}).
For a restriction $G := (S_1, \LL, S_n)$ of $H$ we define
\[
{\cal B}_{i} \stackrel{.}{\cap} G := S_{-i}.
\]

We call then $({\cal B}, \stackrel{.}{\cap})$ the
\oldbfe{pure} belief structure in $H$.
This belief structure was considered in \cite{Ber84}.

A specific case with a different definition of
$\stackrel{.}{\cap}$
was considered in \cite{Pea84}.
In that paper $H$ is a mixed extension of a finite game. So
given initial finite sets of strategies $I_1, \LL, I_n$
each set $T_i$ equals $\Delta I_i$, i.e.,
$H := (\Delta I_1, \LL, \Delta I_n, p_1, \LL, p_n)$.
Then for a restriction $G := (S_1, \LL, S_n)$ of $H$

\[
{\cal B}_{i} \stackrel{.}{\cap} G := \Pi_{j \neq i} \overline{S_{j}},
\]
where for a set $S_j$ of mixed strategies of player $j$,
$\overline{S_{j}}$ denotes its convex hull.

\item Assume $H$ is finite.  ${\cal B}_{i} := \Pi_{j \neq i} \Delta
  T_{j}$ for $i \in [1..n]$.

So beliefs are joint mixed strategies of the opponents.
For a restriction $G := (S_1, \LL, S_n)$ of $H$ we define

\[
{\cal B}_{i} \stackrel{.}{\cap} G := \Pi_{j \neq i} \Delta  S_{j}.
\]

This belief structure was considered in \cite{Ber84}.

\item 
Assume $H$ is finite.
${\cal B}_{i} := \Delta T_{-i}$ for $i \in [1..n]$.

So beliefs are probability distributions over the set of joint pure
strategies of the opponents (usually called \oldbfe{correlated mixed
  strategies}).  For a restriction $G := (S_1, \LL, S_n)$ of $H$ we
define

\[
{\cal B}_{i} \stackrel{.}{\cap} G := \Delta S_{-i}.
\]

This belief structure was mentioned in \cite{Ber84} and studied in
\cite{BD87}, where the term \oldbfe{correlated rationalizability} was
introduced.

\item 
Assume $H$ is finite.  ${\cal B}_{i} := \Pi_{j \neq i} \Delta^{\! \circ}
  T_{j}$ for $i \in [1..n]$.

So beliefs are joint \oldbfe{totally mixed} strategies of the opponents.
For a restriction $G := (S_1, \LL, S_n)$ of $H$ we define

\[
{\cal B}_{i} \stackrel{.}{\cap} G := \Pi_{j \neq i} \Delta^{\! \circ}  S_{j}.
\]

This belief structure was studied in \cite{Pea84}, where a best
response to a belief formed by a joint totally mixed strategy of the
opponents is called a \oldbfe{cautious response}.
\end{enumerate}

Given two finite sets $A$ and $B$ such that $A \sse B$, we identify
each probability distribution on $A$ with the probability distribution
on $B$ in which 0 is assigned to each element in $B \setminus A$.
Then $A \sse B$ implies $\Delta A \sse \Delta B$.
It is now straightforward to see that 
property \textbf{A} is satisfied by the belief structures from
examples (i)---(iii).  So by appropriately choosing the belief
structure $({\cal B}, \stackrel{.}{\cap})$ we shall be able to apply
our results to a variety of frameworks including the ones considered
in \cite{Ber84} and \cite{Pea84}.

It is important, however, to note that property \textbf{A} is not
satisfied by the belief structure from example (iv). The reason is
that for finite sets $A$ and $B$ such that $A \sse B$ the inclusion
$\Delta^{\! \circ} A \sse \Delta^{\! \circ} B$ does not hold.  So the
results of our paper do not apply to the modifications of the notion
of rationalizability that rely on totally mixed strategies, for
example \cite{HV00}, where the notion of weak perfect
rationalizability is studied.\footnote{This is not surprising since by
  the result of \cite{Pea84} the notions of rationalizability
  w.r.t.~totally mixed strategies of the opponent and of not being weakly dominated
  by a mixed strategy coincide in two-player finite games and, as is
  well-known, the latter notion fails to be order independent.}

\section{Operators on complete lattices}
\label{sec:reductions}

We are interested in iterated reductions of strategic games entailed
by various ways of removing strategies.  To deal with them in a
uniform way we define the relevant concepts for arbitrary operators on
a fixed complete lattice $(D, \sse)$ with the largest element $\top$.

In what follows we use ordinals and denote them by $\alpha, \beta, \gamma$.
Given a, possibly transfinite, sequence $(G_{\alpha})_{\alpha < \gamma}$ of
elements of $D$ we denote their join and meet respectively by
$\bigcup_{\alpha < \gamma} G_{\alpha}$
and $\bigcap_{\alpha < \gamma} G_{\alpha}$.

In the subsequent applications $D$ will be the set of all restrictions
of a fixed strategic game $H$ for $n$ players,
ordered by the componentwise set inclusion $\sse$
(so $H$ is the largest element)
and $\bigcup_{\alpha < \gamma}$
and $\bigcap_{\alpha < \gamma}$ the customary set-theoretic 
operations on them. 
But this additional information on the structure of $D$
will be irrelevant in the remainder of this section.

We now establish some general results on operators on complete
lattices.  The proofs are straightforward and the results readily apply
to the operators we shall subsequently study.

\begin{definition}
Let $T$ be an operator on $(D, \sse)$, i.e., $T: D \myra D$.

\begin{itemize}

\item We say that an element $G$ is a \oldbfe{fixpoint} of $T$ if $T(G) = G$.

\item We call $T$ \oldbfe{monotonic} if for all $G_1, G_2$
\[
\mbox{$G_1 \sse G_2$ implies $T(G_1) \sse T(G_2)$.}
\]
\item We call $T$ \oldbfe{contracting} if for all $G$
\[
T(G) \sse G.
\]

\item We define the \oldbfe{contracting version} of $T$ by:
\[
\overline{T}(G) := T(G) \cap G.
\]

\item We define by transfinite induction a sequence of elements
  $T^{\alpha}$ of $D$, where $\alpha$ is an ordinal, as follows:

\begin{itemize}

  \item $T^{0} := \top$,

  \item $T^{\alpha+1} := T(T^{\alpha})$,

  \item for all limit ordinals $\beta$, $T^{\beta} := \bigcap_{\alpha < \beta} T^{\alpha}$.
  \end{itemize}

\item We call the least $\alpha$ such that $T^{\alpha+1} = T^{\alpha}$ the \oldbfe{closure ordinal} of $T$
and denote it by $\alpha_T$.  We call then $T^{\alpha_T}$ the \oldbfe{outcome of} (iterating) $T$.
\HB
\end{itemize}
\end{definition}

The outcome of an operator can be the least element of the 
complete lattice. So in the case of the complete lattice of the
restrictions of $H$ ordered by the componentwise set inclusion $\sse$
the outcome can be an empty restriction.  In general the closure
ordinal of $T$, and hence its outcome, do not need to exist. (Take for
example an operator oscillating between two values.)  However, we have
the following classic result due to \cite{Tar55}.\footnote{We use here
  its `dual' version in which the iterations start at the largest and
  not at the least element of a complete lattice.}  \II

\NI
\textbf{Tarski's Fixpoint Theorem} 
Every monotonic operator on $(D, \sse)$ has a largest fixpoint. This
fixpoint is the outcome of $T$, i.e., it is of the form
$T^{\alpha_T}$. 
\vspace{1mm}

Clearly, if $T$ is monotonic, then so is
$\overline{T}$.
Other observations concerning the above notions are gathered
in the following note.

\begin{note}\label{not:contracting}
Consider an operator $T$ on $(D, \sse)$.
\begin{enumerate} \smallromani
  
\item If $T$ is contracting or monotonic, then $T^{\alpha+1} \sse
  T^{\alpha}$ for all $\alpha$.

\item Suppose that $T^{\alpha+1} \sse T^{\alpha}$
for all $\alpha$.
Then 

\begin{itemize}

\item $\overline{T}^{\alpha} = T^{\alpha}$
for all $\alpha$,

\item the outcomes of $T$ and $\overline{T}$ exist
and coincide.
\end{itemize}
\end{enumerate} 
\end{note}
\Proof
The existence of the outcome of $T$ under the assumption of $(ii)$
follows by the standard arguments
of set theory (see, e.g., \cite{Acz77}).
The other claims follow by transfinite induction.
\HB
\VV

So if we are only interested in the operator iterations or its
outcome, it does not matter whether we choose an operator $T$ that is
contracting or monotonic, or its contracting version $\overline{T}$.

In what follows we shall frequently use
the following lemma.

\begin{lemma} \label{lem:inc}
Consider two operators $T$ and $R$ on $(D, \sse)$ such that
\begin{itemize}
\item for all $\alpha$, $T(R^{\alpha}) \sse R(R^{\alpha})$,

\item $T$ is monotonic.

\end{itemize}
Then for all $\alpha$
\[
T^{\alpha} \sse R^{\alpha}.
\]

In particular if $R$ has an outcome, then
$T^{\alpha_{T}} \sse R^{\alpha_{R}}$.

\end{lemma}

\Proof
We prove the first claim by transfinite induction.
By the definition of the iterations we only need to consider the induction
step for a successor ordinal.  So suppose the claim holds for some
$\alpha$. Then by the first two assumptions and the induction
hypothesis we have the following string of inclusions and equalities:
\[
T^{\alpha + 1} =   T(T^{\alpha}) \sse T(R^{\alpha}) \sse R(R^{\alpha}) = R^{\alpha + 1}.
\]

By Tarski's Fixpoint Theorem $T$ has an outcome, so the second claim 
follows immediately from the first.
\HB
\VV

We now formalize the idea that a procedure of iterated elimination of
strategies is order independent. Interestingly, it is possible to
state this property without specializing the underlying complete
lattice to that of all restrictions of the initial game, by 
simply focusing on iterations of operators. The
definition we provide also takes care of the possibility that the
elimination process takes more than $\omega$ iteration steps.

\begin{definition} \label{def:oi1}
  Consider a contracting operator $T$ ('$T$ removes strategies') on
  $(D, \sse)$. We say that $T$ is \oldbfe{order independent} if
\[
R^{\alpha_R} = T^{\alpha_T}
\]
('the outcomes of the iterated eliminations of strategies coincide')
for each operator $R$ such that for all $\alpha$

\begin{itemize}
\item $T(R^{\alpha}) \sse R(R^{\alpha}) \sse R^{\alpha}$

('$R$ removes from $R^{\alpha}$ only strategies that $T$ removes')

\item if $T(R^{\alpha}) \subsetneq R^{\alpha}$, then
$R(R^{\alpha}) \subsetneq R^{\alpha}$

('if $T$ can remove some strategies from $R^{\alpha}$,
then $R$ as well').
\end{itemize}
(Note that $R^{\alpha_R}$ exists by Note
\ref{not:contracting}.)
\HB
\end{definition}

The intuitions provided in the brackets hopefully make this definition
self-explanatory.  This definition is perfectly satisfactory as long
as we consider procedures that \emph{remove} strategies. However, 
we shall consider here arbitrary procedures on games and
sometimes they may \emph{add} strategies.
To deal with this more general setting we introduce a definition of
order independence for arbitrary operators.

\begin{definition} \label{def:oi2}
Consider an operator $T$ on $(D, \sse)$.

\begin{itemize}
\item 
We say that the operator $R$ is a \oldbfe{relaxation} of $T$ if for all $\alpha$
\begin{itemize}

\item $T(R^{\alpha}) \sse R(R^{\alpha})$,

\item if $T(R^{\alpha}) \sse R^{\alpha}$, then $R(R^{\alpha}) \sse R^{\alpha}$,

\item if $R^{\alpha}$ is a fixpoint of $R$, then it is a fixpoint of $T$.

\end{itemize}

\item We say that $T$ is \oldbfe{order independent} if
the set
\[
\mbox{\{$G \mid G$ is an outcome of a relaxation of $T$\}}
\]
has at most one element.
\HB
\end{itemize}
\end{definition}

The intuition behind the conditions defining a relaxation is as follows.  Suppose that 
$T$ is a procedure on the set of restrictions of the initial game.
Then the first condition states that during the iterations of $R$ (i.e., for
all restrictions $R^{\alpha}$) the operator $R$
`approximates' $T$ from above.  In turn, the second condition states
that if $T$ removes some strategies from some restriction $R^{\alpha}$, then so does $R$.  
Finally, the third condition states
that if some strategy can be removed from/added to some restriction $R^{\alpha}$ by the operator
$T$ ($R^{\alpha}$ is not a fixpoint of $T$), then the same holds for the $R$ operator
($R^{\alpha}$ is not a fixpoint of $R$).  

Note that this definition of order independence of an operator $T$ does
not even require that $T$ has an outcome. If it does have, then each
relaxation of it has the same outcome as $T$, if it has one.

The following observation shows that the second definition
generalizes the first one.

\begin{note}
For a contracting operator $T$ on $(D, \sse)$ 
both definitions of order independence coincide.
\end{note}
\Proof
For a contracting operator $T$ the first two conditions of the definition
of a relaxation $R$ are equivalent to the first condition on $R$ in 
Definition \ref{def:oi1}, while the third condition is equivalent
to the second condition on $R$ in 
Definition \ref{def:oi1}.

Moreover, for all relaxations $R$ of $T$ we have $R^{\alpha+1} \sse
R^{\alpha}$ for all $\alpha$ and hence, by
Note \ref{not:contracting}, they all have an outcome. Also $T$ is a
relaxation of itself, so order independence in the sense of Definition
\ref{def:oi2} is equivalent to the statement that $R^{\alpha_R} =
T^{\alpha_T}$ holds for all relaxations $R$ of $T$.  
\HB 
\VV

From now on when referring to order independence we shall mean the 
latter definition. The following general result holds.

\begin{theorem}[Order Independence] \label{thm:order}
Every monotonic operator on $(D, \sse)$ is order independent.
\end{theorem}

\Proof Let $T$ be a monotonic operator.  By Tarski's Fixpoint Theorem
$T$ has an outcome, $T^{\alpha_T}$, which is the largest fixpoint of $T$.
To prove the theorem it suffices to show that
\[
\mbox{\{$G \mid G$ is an outcome of a relaxation of $T$\} = \{$T^{\alpha_T}$\}.}
\]

So take a relaxation $R$ of $T$ that has an outcome, $R^{\alpha_R}$.
By Lemma \ref{lem:inc} $T^{\alpha_T} \sse R^{\alpha_R}$.
But by the definition of a relaxation
$R^{\alpha_R}$ is a fixpoint of $T$ and $T^{\alpha_T}$ is the
largest fixpoint of $T$, so also $R^{\alpha_R} \sse
T^{\alpha_T}$. Consequently $R^{\alpha_R} = T^{\alpha_T}$.  
\HB 
\VV

Intuitively, this result can be interpreted as follows.  Suppose that
an operator $T$ removes \emph{all} strategies from a
restriction of a game $H$ that meet some property and that we iterate this
operator starting with $H$.  Suppose now that at each stage we remove
only \emph{some} strategies that meet this property (instead of
all). Then, when $T$ is monotonic, we
still end up with the same outcome (which is the largest
fixpoint of the $T$ operator).

Note that we did not assume here that the operator $T$ is contracting
and consequently had to rely on the second definition of order
independence. 
Also, in the proof we did not use the second condition
of the definition of a relaxation.  In our approach this condition is
needed only to deal with the contracting operators that are not
monotonic.




\section{The $\emph{\GR}$ and $\overline{\emph{\GR}}$ operators}
\label{sec:st}

We now apply the above considerations to specific operators.
Each of them is defined in the context of a fixed game $H$ and 
a belief structure $({\cal B}, \stackrel{.}{\cap})$ in $H$. 

Given a restriction $G := (S_1, \LL, S_n)$ of $H$ and a belief
$\mu_i \in {\cal B}_{i} \stackrel{.}{\cap} G$
we say that a strategy $s_i$ of player $i$ in the game $H$
is a \oldbfe{best
  response to} $\mu_i$ \oldbfe{in} $G$, and
write $s_i \in BR_G(\mu_i)$, if 
\[
\fa s'_i \in S_i \: p_i(s_i, \mu_i) \geq p_i(s'_i, \mu_i).
\]
Note that $s_i$ does not need to be an element of $S_i$.

We now introduce the following operator $\textit{\GR}$ (standing for
`global rationalizability') on the set of restrictions of $H$:

\[
\GR(G) := (S'_1, \LL, S'_n),
\]
where for all $i \in [1..n]$

\[
S'_i := \{ s_i \in T_i \mid \te \mu_{i} \in {\cal B}_{i} \stackrel{.}{\cap} G \: s_i \in BR_H(\mu_i)\}.
\]

So $\GR(G)$ is obtained by removing from $H$ all strategies that are
never best responses \emph{in} $H$ (note this reference to $H$) to a
belief held in $G$. That is, when removing the strategies we allow
justifications (for their removal) from the initial game $H$.  

Thanks to property \textbf{A} the operator $\GR$ is monotonic. So we
can apply to $\GR$ and its contracting version $\overline{\GR}$ Note
\ref{not:contracting} and as a result we can confine further analysis to the
latter operator.  The $\GR$ operator was considered in \cite{Ber84}
(called there $\lambda$, see page 1015).\footnote{The reader may now notice that in
Example \ref{exa:Bertrand1} we used the $\overline{\GR}$ operator instead of $\GR$, 
which is more complicated to justify informally.}
On the account of the results from the previous section the $\overline{\GR}$ operator enjoys various properties.
We list them as the following result.

\begin{theorem}[$\overline{\GR}$] \label{thm:st}
\mbox{} \hspace{-8mm}

\begin{enumerate} \smallromani

\item The largest fixpoint of $\overline{\GR}$ exists and is its outcome.

\item $\overline{\GR}$ is order independent.
  
\item For all relaxations $R$ of $\overline{\GR}$ and all $\alpha$ we have
  $\overline{\GR}^{\alpha} \sse R^{\alpha}$.
\HB
\end{enumerate}
\end{theorem}

In general, infinite iterations of $\overline{\GR}$ can be necessary. In fact,
in some games $\omega$ iterations of $\overline{\GR}$
are insufficient to reach the outcome, that is, the closure ordinal of $\overline{\GR}$ can be larger than $\omega$.

\begin{example} \label{exa:4}
Consider the following game $H$ with two players.
The set of strategies for each player is the set of natural numbers ${\cal N}$
augmented by $-1$.
The payoff functions are defined as follows.
For $k, \ell \geq 0$ we put:

\[
\begin{array}{l}
p_1(k, \ell) := \left\{ 
\begin{tabular}{ll}
$\ell + 1$ &  \mbox{if $k = \ell +1$} \\
0 &  \mbox{otherwise}
\end{tabular}
\right . \\
\\
p_2(k, \ell) := \left\{ 
\begin{tabular}{ll}
$k$ &  \mbox{if $k = \ell$} \\
0 &  \mbox{otherwise}
\end{tabular}
\right . 
\end{array}
\]
For the remaining pairs of strategies we put
for $k, \ell \geq -1$ and $k_0, \ell_0 \geq 0$:
\[
\begin{array}{l}
p_1(-1, \ell) := \ell +1, \\
p_1(k_0, -1) := k_0, \\
p_2(k, -1) := k, \\
p_2(-1, \ell_0) := \ell_0. \\
\end{array}
\]

Further we assume the pure belief structure, i.e., the beliefs are the
strategies of the opponent.  The following two diagrams explain the
structure of this game.  In each of them on the left we depict
strategies of player 1 and on the right the strategies of player 2. An
arrow from $a$ to $b$ stands for the statement `strategy $a$ is a best
response to strategy $b$'.  In particular, no arrow leaves strategy 0
of player 1, which indicates that it is not a best response to any
strategy of player 2.

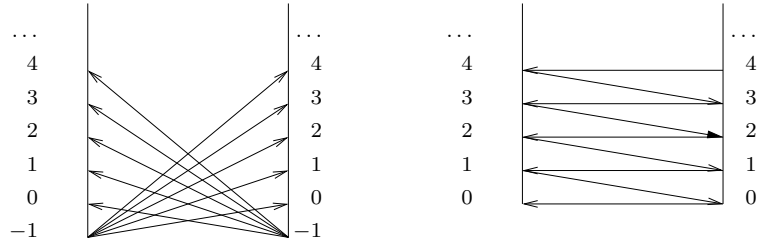
\begin{figure}[htbp]
  \centering
\input{minusone.pstex_t}   
  \label{fig:game}
\end{figure}

Note that 

\begin{itemize}
  
\item each $-1$ is a best response to any non-negative strategy,

\item no best response to any $-1$ strategy exists,

\item the best response to strategy $k \geq 0$ of player 1 is $k$,

\item the best response to strategy $\ell \geq 0$ of player 2 is $\ell + 1$.
\end{itemize}

Denote ${\cal N} \cup \{-1\}$ by ${\cal N}'$. 
It is easy to see that 
\[
 \begin{array}{l}
\overline{\GR}^{0} = ({\cal N}', {\cal N}'), \\
\overline{\GR}^{1} = ({\cal N}' \setminus \C{0}, {\cal N}'), \\
\overline{\GR}^{2} = ({\cal N}' \setminus \C{0}, {\cal N}' \setminus \C{0}), \\
\overline{\GR}^{3} = ({\cal N}' \setminus \C{0,1}, {\cal N}' \setminus \C{0}), \\
\overline{\GR}^{4} = ({\cal N}' \setminus \C{0,1}, {\cal N}' \setminus \C{0,1}), \\
 \LL
\end{array}
\]

So $\overline{\GR}^{\omega} = (\{-1\}, \{-1\})$.
But in the initial game no $-1$ is a best response to `the other' $-1$. So
$\overline{\GR}((\{-1\}, \{-1\})) = (\ES, \ES)$, that is $\overline{\GR}^{\omega + 1} = (\ES, \ES)$.
\HB
\end{example}

As already mentioned in Section \ref{sec:introduction}, in
\cite{Lip94} a two-player game is constructed for which $\omega$
iterations of $\GR$ are insufficient to reach the outcome.  In this
example each set of beliefs ${\cal B}_i$ consists of the mixed
strategies of the opponent and the game is considerably more complex.

\section{The $\textit{\LR}$ and $\overline{\textit{\LR}}$ operators}
\label{sec:rat}

In this section we analyze 
the following operator $\textit{\LR}$ 
(standing for `local rationalizability')
on the set of restrictions of $H$:

\[
\textit{\LR}(G) := (S'_1, \LL, S'_n),
\]
where for all $i \in [1..n]$

\[
S'_i := \{ s_i \in T_i \mid \te \mu_{i} \in {\cal B}_{i} \stackrel{.}{\cap} G \: s_i \in BR_G(\mu_i)\}.
\]

So $\textit{\LR}(G)$ is obtained by removing from $H$ all strategies that are
never best responses \emph{in} $G$ (so \emph{not} in $H$) to a belief
held in $G$. That is, when removing strategies we use justifications
(for their removal) from the currently considered game $G$.  
For each restriction $G$ of $H$, $s_i \in BR_H(\mu_i)$ implies $s_i \in BR_G(\mu_i)$, 
so for all restrictions $G$ we have $\GR(G) \sse \textit{\LR}(G)$.

Unfortunately, the $\textit{\LR}$ operator cannot be used as a
meaningful basis for the rationalizability notion.  

\begin{example}
To see this reconsider
the game from Example \ref{exa:4}.
We first prove by induction that for $n \geq 0$ we have
$\textit{\LR}^{n} = \overline{\GR}^{n}$. 

The base case obviously holds. Denote the set $\C{0, \LL, m}$ by $I_m$.
For the induction step note that if
$n = 2k$, then
\[
\textit{\LR}^{2k} = \overline{\GR}^{2k} = 
({\cal N}' \setminus I_k, {\cal N}' \setminus I_k).
\]
For any strategy $m > k$ we have both
\[
p_1(m+1, m) = m+1 > 0 = p_1(k+1, m)
\]
and
\[
p_1(m+1, -1) = m+1 > k+1 = p_1(k+1,-1).
\]
This shows that for each strategy $\ell \in {\cal N}' \setminus I_k$
of player 2 the strategy $k+1$ of player 1 is not a best response to
$\ell$ in $\textit{\LR}^{2k}$ and consequently
\[
\textit{\LR}^{2k+1} = 
({\cal N}' \setminus I_{k+1}, {\cal N}' \setminus I_k) = 
\overline{\GR}^{2k+1}.
\]
A similar argument applies when $n = 2k+1$.  

Consequently $\textit{\LR}^{\omega} = \textit{\GR}^{\omega} = (\{-1\},
\{-1\})$.  But for all $s_1, s_2 \in {\cal N}'$ we have $p_1(s_1, -1)
\geq p_1(-1, -1) = 0$ and $p_2(-1, s_2) \geq p_2(-1, -1) = -1$. So for
each player each strategy from ${\cal N}'$ is a best response to the
(unique) strategy $-1$ of the other player in $\textit{\LR}^{\omega}$.
Hence $\textit{\LR}^{\omega+1} = ({\cal N}', {\cal N}')$.  This shows
that \textit{\LR} has no outcome.  Also it is not monotonic and not
even contracting.
\HB
\end{example}



Consequently, we confine our attention to its contracting version, the
$\overline{\textit{\LR}}$ operator (defined by
$\overline{\textit{\LR}}(G) := \textit{\LR}(G) \cap G$), that always has
an outcome.  Note for example that for the game from Example
\ref{exa:4} we have $\overline{\LR}^{\omega} = (\{1\}, \{1\})=
\overline{\LR}^{\omega+1}$, so $(\{-1\}, \{-1\})$ is the outcome of
$\overline{\textit{\LR}}$.  This operator was introduced implicitly and
for specific games, in \cite{Pea84}, in Definition 1 on page 1032.

Unfortunately, $\overline{\textit{\LR}}$ fails to satisfy any of the
properties of $\overline{\GR}$ listed in the $\overline{\GR}$ Theorem
\ref{thm:st}.  First, the largest fixpoint of
$\overline{\textit{\LR}}$ does not need to exist.  Indeed, let us
return to the game from Example \ref{exa:4}.  For all $x \in {\cal N}
\cup \{-1\}$ the restriction $(\{x\}, \{x\})$ is a fixpoint of
$\overline{\textit{\LR}}$. However, their union, $H$, is not.
Consequently, by Tarski's Fixpoint Theorem $\overline{\textit{\LR}}$
is not monotonic.




Next, we exhibit a relaxation $R$ of $\overline{\textit{\LR}}$ the outcome of which
is different than the outcome of $\overline{\textit{\LR}}$.

\begin{example} \label{exa:1}
  Consider a two-player game $H$ in which the set of strategies for
  each player is the set ${\cal N}$ of natural numbers. The payoff to
  each player is the number (strategy) he selected. Take as the set of
  mixed strategies the probability distributions on ${\cal N}$ with a
  finite support (sometimes called simple probabilities).  So when
  computing the payoffs to mixed strategies each of them can be
  written as $\sum_{i \in A} \pi(i) \cdot i$, where $A$ is a finite
  subset of ${\cal N}$ and $\pi$ is a probability distribution on $A$.
  (We shall return to this definition in Section \ref{sec:mixed}.)
  Take the belief structure in which the beliefs of each player is the
  set of so defined mixed strategies of the opponent.

Clearly, no strategy is a best response to a mixed strategy of
  the opponent. So $\overline{\textit{\LR}}(H) = (\ES, \ES)$ and $(\ES, \ES)$ is
  the outcome of $\overline{\textit{\LR}}$.  However, for any $i \geq
  0$ the restriction $(\{i\}, \{i\})$ is the outcome of a relaxation
  $R$ of $\overline{\textit{\LR}}$ such that $R(H) = (\{i\}, \{i\})$.
  \HB
\end{example}

This example shows that the rationalizability notion entailed by the
$\overline{\textit{\LR}}$ operator is not order independent.  In this example the outcome
of a relaxation of $\overline{\textit{\LR}}$ is strictly larger than the outcome of $\overline{\textit{\LR}}$.
A more troublesome example is the following one in which the
outcome of a relaxation of $\overline{\textit{\LR}}$ is strictly smaller than the outcome
of $\overline{\textit{\LR}}$.

\begin{example} \label{exa:Bertrand}
We return here to Example \ref{exa:Bertrand1} from Section \ref{sec:introduction}.
So $H$ is a game with two players, each with the set $(0, 100]$ of strategies,
and the payoff functions are defined by:

\[
\begin{array}{l}
p_1(s_1, s_2) := \left\{ 
\begin{tabular}{ll}
$s_1 (100 - s_1) $ &  \mbox{if $s_1 < s_2$} \\[2mm]
$\dfrac{s_1 (100 - s_1)}{2} $ &  \mbox{if $s_1 = s_2$} \\[2mm]
0 &  \mbox{if $s_1 > s_2$} 
\end{tabular}
\right .  \\
\\
p_2(s_1, s_2) := \left\{ 
\begin{tabular}{ll}
$s_2 (100 - s_2) $ &  \mbox{if $s_2 < s_1$} \\[2mm]
$\dfrac{s_2 (100 - s_2)}{2} $ &  \mbox{if $s_1 = s_2$} \\[2mm]
0 &  \mbox{if $s_2 > s_1$} 
\end{tabular}
\right . 
\end{array}
\]
Also, we assume the pure belief structure.  

We noted already in Example \ref{exa:Bertrand1} that
$\overline{\textit{\LR}}(H) = (\C{50}, \C{50})$ and \\
$\overline{\textit{\LR}}((\C{50}, \C{50}))
= ((\C{50}, \C{50}))$.

Take now the relaxation $R$ of $\overline{\textit{\LR}}$ defined by:
\[
R(G):= \left\{ 
\begin{tabular}{ll}
((0, 50], (0, 50]) &  \mbox{if $G = H$} \\
\mbox{$\overline{\textit{\LR}}(G)$} &  \mbox{otherwise}
\end{tabular}
\right . 
\]

Then $R(H) = (0, 50], (0, 50])$ and 
$R(((0, 50], (0, 50])) = (\ES, \ES)$ since \\
$\overline{\textit{\LR}}(((0, 50], (0, 50])) = (\ES, \ES)$.
This shows that $H$ cannot be reduced to an empty restriction by $\overline{\textit{\LR}}$ though it can by 
some relaxation of it.

Finally, note that in this example also the 
$\mathit{\LR}$ operator exhibits an erratic behaviour. Indeed, 
$\textit{\LR}(H) = (\C{50}, \C{50})$, i.e., $\textit{\LR}^1 = (\C{50}, \C{50})$.
However, it is easy to see that $(49, 49) \in \textit{\LR}(\C{50}, \C{50})$, 
since $p_1(49, 50) > p_1(50, 50)$ and $p_2(50, 49) > p_2(50, 50)$.
So $\textit{\LR}^2$ is not a restriction of $\textit{\LR}^1$.
\HB
\end{example}

This example suggest that $\overline{\textit{\LR}}$ is unnecessarily
weak.  Also, it is counterintuitive that when
$\overline{\textit{\LR}}$ is used to define rationalizability, in some
natural games it is beneficial to eliminate at certain stages only
\emph{some} strategies that are never best responses.

It is also easy to see that in Example \ref{exa:Bertrand}
the outcome
of the $\overline{\GR}$ operator is an empty restriction.  Indeed, $\overline{\GR}(H) =
(\C{50}, \C{50})$ and $\overline{\GR}((\C{50}, \C{50})) = (\ES, \ES)$, since $s_1 = 49$
is a better response in $H$ to $s_2 = 50$ than $s_1 = 50$ and
symmetrically for the second player.
So $\overline{\GR}$ is strictly stronger than $\overline{\textit{\LR}}$ in the sense that its outcome 
can be strictly smaller than that of $\overline{\textit{\LR}}$.

In this example the outcome of $\overline{\GR}$ is the same as the outcome of
a relaxation of $\overline{\textit{\LR}}$.  However, this does not need to hold in general.
In other words, $\overline{\GR}$ defines a genuinely more powerful rationalizability notion
that cannot be derived from $\overline{\textit{\LR}}$.  

Indeed, reconsider the game from Example \ref{exa:4}.  It is easy to
see that the finite iterations of $\overline{\textit{\LR}}$ and
$\overline{\GR}$ coincide, so $\overline{\textit{\LR}}^{\omega} =
(\{-1\}, \{-1\})$.  But $(\{-1\}, \{-1\})$ is a fixpoint of
$\overline{\textit{\LR}}$, so using $\overline{\textit{\LR}}$ we cannot
reduce the initial game to an empty restriction.  Moreover, the same
holds for any relaxation $R$ of $\overline{\textit{\LR}}$. Indeed, it
is easy to see that for each relaxation $R$ of
$\overline{\textit{\LR}}$ its outcome always contains the joint
strategy $(-1, -1)$.

\section{When the outcomes of $\overline{\textit{\GR}}$ and $\overline{\textit{\LR}}$ coincide}
\label{sec:coincide}

To analyze the situations when the iterations of the $\overline{\GR}$ and
$\overline{\textit{\LR}}$ operators coincide we introduce the following
property of the initial game $H$, given a belief structure 
$({\cal B}, \stackrel{.}{\cap})$ in $H$.

\begin{description}

\item[B] For all beliefs $\mu_i \in {\cal B}_{i}$
a best response to $\mu_i$ in $H$ exists.
\end{description}

The importance of this property stems from the fact that it allows us
to equate during each iteration of the $\overline{\textit{\LR}}$
operator the best responses in the `current' game with the best
responses in the initial game.

For the finite games and all belief structures property \textbf{B}
obviously holds but it can clearly fail for the infinite games. For
instance, it does not hold in the game considered in Example
\ref{exa:1} since in this game no strategy is a best response to a
strategy of the opponent.  Also it does not hold in the game from
Example \ref{exa:4} but we now offer an example of a game in
which property \textbf{B} does hold and in which the iterations of length
$> \omega$ of both $\overline{\GR}$ and $\overline{\textit{\LR}}$ 
are still unavoidable.

\begin{example} \label{exa:2}
We modify the game from Example \ref{exa:4} by adding the third player and by 
removing the strategy $-1$ from the set of strategies of each player.
So the set of strategies for each player is the set of natural numbers ${\cal N}$.
The payoff functions are now defined as follows:
\[
\begin{array}{l}
p_1(k, \ell, m) := \left\{ 
\begin{tabular}{ll}
$\ell + 1$ &  \mbox{if $k = \ell +1$} \\
0 &  \mbox{otherwise}
\end{tabular}
\right . \\
\\
p_2(k, \ell, m) := \left\{ 
\begin{tabular}{ll}
$k$ &  \mbox{if $k = \ell$} \\
0 &  \mbox{otherwise}
\end{tabular}
\right .  \\
\\
p_3(k, \ell, m) := 0.
\end{array}
\]

Further we assume the pure belief structure.
Note that

\begin{itemize}
\item the best response to $s_{-1} = (\ell, m)$ is $\ell +1$,

\item the best response to $s_{-2} = (k, m)$ is $k$,

\item each $m \in {\cal N}$ is a best response to $s_{-3} = (k, \ell)$.

\end{itemize}

So to each joint strategy of the opponents a best response exists. That is, 
property \textbf{B} is satisfied.
Analogously to Example \ref{exa:4} we have:
\[
\begin{array}{l}
\overline{\GR}^{0} = ({\cal N}, {\cal N}, {\cal N}), \\
\overline{\GR}^{1} = ({\cal N} \setminus \C{0}, {\cal N}, {\cal N}), \\
\overline{\GR}^{2} = ({\cal N} \setminus \C{0}, {\cal N} \setminus \C{0}, {\cal N}), \\
\overline{\GR}^{3} = ({\cal N} \setminus \C{0,1}, {\cal N} \setminus \C{0}, {\cal N}), \\
\overline{\GR}^{4} = ({\cal N} \setminus \C{0,1}, {\cal N} \setminus \C{0,1}, {\cal N}), \\
\LL
\end{array}
\]

So $\overline{\GR}^{\omega} = (\ES, \ES, {\cal N})$.
Also $\overline{\GR}((\ES, \ES, {\cal N})) =  (\ES, \ES, \ES)$, so 
$\overline{\GR}^{\omega + 1}  = (\ES, \ES, \ES)$.

Finally, it is easy to see that $\overline{\textit{\LR}}^{\alpha} =
\overline{\GR}^{\alpha}$ for all $\alpha$ (this is in fact, a
consequence of a result proved below), so $\omega +1$ is the closure
ordinal of $\overline{\textit{\LR}}$, as well.  \HB
\end{example}

For infinite games a natural situation when property \textbf{B} holds
is the following.  Recall that a game $H = (T_1, \LL, T_n, p_1, \LL,
p_n)$ is called \oldbfe{compact} if the strategy sets are 
compact subsets of a complete metric space, and
\oldbfe{own-uppersemicontinuous} if each payoff function $p_i$ is
uppersemicontinuous in the $i$th argument. In turn, $p_i$ is called
\oldbfe{uppersemicontinuous in the} $i$\oldbfe{th argument} if the set
$\C{s'_i \in T_i \mid p_i(s'_i, s_{-i}) \geq r}$ is closed for all $r
\in {\cal R}$ and all $s_{-i} \in T_{-i}$.
  
As explained in \cite{DS02} (see the proof of Lemma on page 2012) for
such games and pure belief structures property \textbf{B} holds by
virtue of a standard result from topology.  If we impose a stronger
condition on the payoff functions, namely that each of them is
continuous ---the game is called then \oldbfe{continuous}--- then we
are within the framework considered in \cite{Ber84}.  As shown there
for compact and continuous games iterations of length $> \omega$ for
the $\GR$ operator do not need to be considered, that is, its outcome
(or equivalently the outcome of $\overline{\GR}$) can be reached in
$\omega$ iterations.  To put it more succinctly: $\alpha_{\GR} \leq
\omega$.

In the presence of properties \textbf{A} and \textbf{B} still many
differences between the operators $\overline{\GR}$ and
$\overline{\textit{\LR}}$ remain.  In particular,
$\overline{\textit{\LR}}$ does not need to be monotonic and its largest
fixpoint does not need to exist.  Indeed, consider the following example
in which we consider a finite game, which ensures
property \textbf{B}.

\begin{example} \label{exa:012}
Consider the game $H := (A, A, p_1, p_2)$,
where $A = \{1, \LL, n\}$ and 
for $i,j \in A$
\[
p_1(i, j) := i,
\]
\[
p_2(i, j) := 1.
\]
Assume the belief structure in which the beliefs of each player is the set of
the mixed strategies of the opponent, i.e., ${\cal B}_{1} = {\cal B}_{2} = \Delta A$.

Note that

\begin{itemize}

\item $n$ is the unique best response of player 1 to any belief $\mu_1 \in {\cal B}_{1}$ about player 2,

\item each $j \in A$ is a best response of player 2 to any belief $\mu_2 \in {\cal B}_{2}$ about player 1.

\end{itemize}
So $\overline{\textit{\LR}}(H)= (\{n\}, A)$ and $ (\{n\}, A)$ is a fixpoint of $\overline{\textit{\LR}}$.
Also each restriction $(\{j\}, \{j\})$ for $j \in A$ is a fixpoint of $\overline{\textit{\LR}}$. However,
their union, $H$, is not.

Furthermore $(\{1\}, \{1\}) \sse H$, but $\overline{\textit{\LR}}(\{1\}, \{1\}) = (\{1\}, \{1\})$ is not a restriction 
of $\overline{\textit{\LR}}(H) = (\{n\}, A)$, so $\overline{\textit{\LR}}$ is not monotonic.
\HB
\end{example}

To prove the positive results we identify first the crucial property
of the initial game $H = (T_1, \LL, T_n, p_1, \LL,p_n)$ that holds in
the presence of property \textbf{B}.

\begin{lemma}\label{lem:b}
Assume property \textbf{B}. Then 
for all  relaxations $R$ of $\overline{\textit{\LR}}$ and all $i \in [1..n]$
\[
\fa \alpha \: \fa \mu_i  \in {\cal B}_{i} \stackrel{.}{\cap}  R^{\alpha} \: \fa s_i \in T_i  \: 
(s_i \in BR_{R^{\alpha}}(\mu_i) \myra s_i \in BR_{H}(\mu_i)),
\]
where $R^{\alpha} := (S_1, \LL, S_n)$.
\end{lemma}

In words, for all $\alpha$ and all beliefs in 
$\mu_i \in {\cal B}_{i} \stackrel{.}{\cap}  R^{\alpha}$ if $s_i$ is
a best response to $\mu_i$ in $R^{\alpha}$, then it is in fact
a best response to $\mu_i$ in $H$.
\II

\Proof Suppose $s_i \in BR_{R^{\alpha}}(\mu_i)$, where $\mu_i \in
{\cal B}_{i} \stackrel{.}{\cap} R^{\alpha}$. The operator
$\overline{\textit{\LR}}$ is contracting, so by the second condition in the
definition of a relaxation for all $\beta \leq \alpha$ we have
$R^{\alpha} \sse R^{\beta}$.  Therefore by property \textbf{A} for all
$\beta \leq \alpha$ we have $\mu_{i} \in {\cal B}_{i}
\stackrel{.}{\cap} R^{\beta}$.  In particular $\mu_{i} \in {\cal
  B}_{i}$.  

By property \textbf{B} a best response $s^{*}_i$ to
$\mu_{i}$ in $H$ exists.  But $s^{*}_i \in BR_H(\mu_i)$ implies that
for all $\beta \leq \alpha$, $s^{*}_i \in BR_{R^{\beta}}(\mu_i)$, so 
for all $\beta \leq \alpha$, 
$s^{*}_i \in \overline{\textit{\LR}}(R^{\beta}) \sse R^{\beta+1}$ (the
inclusion holds since $R$ is a relaxation of
$\overline{\textit{\LR}}$).
Moreover $s^{*}_i \in R^{0} = H$ and if $s^{*}_i \in R^{\beta}$
for all successor ordinals $\beta \leq \alpha$, then also 
$s^{*}_i \in R^{\beta}$
for all limit ordinals $\beta \leq \alpha$.
So $s^{*}_i$ is a strategy of player $i$
in all restrictions $R^{\beta}$ for $\beta \leq \alpha$, in particular
$R^{\alpha}$. Hence
\[
p_{i}(s^{*}_i, \mu_{i}) = p_{i}(s_i, \mu_{i}).
\]
So $s_i $ is also a best response to $\mu_{i}$ in $H$. 
\HB
\VV

We prove now the following positive results under the assumption of property \textbf{B}.
Part $(i)$ shows that arbitrary iterations of the $\overline{\GR}$ and
$\overline{\LR}$ operators coincide.
Part $(ii)$ establishes order independence of the operator
$\overline{\LR}$.  This is not an immediate consequence of part $(i)$
since in general the relaxations of $\overline{\GR}$ and
$\overline{\LR}$ differ.  Finally, in part $(iii)$ we return to the
$\textit{\LR}$ operator that we considered and rejected in the previous
section.

\begin{theorem}\label{thm:b}
Assume properties \textbf{A} and \textbf{B}.

\begin{enumerate} \smallromani
\item For all $\alpha$ we have $\overline{\GR}^{\alpha} = \overline{\textit{\LR}}^{\alpha}$.
In particular $\overline{\GR}^{\alpha_{\overline{\GR}}} = \overline{\textit{\LR}}^{\alpha_{\overline{\textit{\LR}}}}$.

\item The $\overline{\LR}$ operator is order independent.

\item For all $\alpha$ we have $\textit{\LR}^{\alpha} = \overline{\textit{\LR}}^{\alpha}$.
In particular $\textit{\LR}^{\alpha_{\textit{\LR}}}$
exists and
$\textit{\LR}^{\alpha_{\textit{\LR}}} = 
\overline{\textit{\LR}}^{\alpha_{\overline{\textit{\LR}}}}$.
\end{enumerate}
\end{theorem}

\Proof 

\NI
$(i)$
For all $G$ we have $\overline{\GR}(G) \sse \overline{\LR}(G)$, so
in view of Lemma \ref{lem:inc} we only need to prove for all $\alpha$ the
inclusion $\overline{\textit{\LR}}^{\alpha} \sse
\overline{\GR}^{\alpha}$. 
We proceed by transfinite induction.
Suppose that $\overline{\textit{\LR}}^{\alpha} \sse
\overline{\GR}^{\alpha}$.

Let $s_i$ be a strategy of player $i$ in $\overline{\textit{\LR}}^{\alpha+1}$.
By definition for some $\mu_{i} \in {\cal B}_{i} \stackrel{.}{\cap}  \overline{\textit{\LR}}^{\alpha}$
we have $s_i \in BR_{\overline{\textit{\LR}}^{\alpha}}(\mu_i)$, so by Lemma \ref{lem:b}
$s_i \in BR_{H}(\mu_i)$. 
Further, $\overline{\textit{\LR}}^{\alpha+1} \sse \overline{\textit{\LR}}^{\alpha}$, so by
the induction hypothesis $s_i$ is a strategy of player $i$ in
$\overline{\GR}^{\alpha}$. So $s_i$ is a strategy of player $i$ in $\overline{\GR}^{\alpha+1}$.

The induction step for a limit ordinal is immediate.
\II

\NI
$(ii)$
Consider a relaxation $R$ of $\overline{\textit{\LR}}$ that has an outcome.
$R^{\alpha_R}$ is a fixpoint of $R$, so
$R^{\alpha_R}$ is also a fixpoint of $\overline{\textit{\LR}}$.
Let $s_i$ be a strategy of player $i$ in $R^{\alpha_R}$.
By the definition of $\overline{\textit{\LR}}$
for some $\mu_{i} \in {\cal B}_{i} \stackrel{.}{\cap}  R^{\alpha_R}$
we have $s_i \in BR_{R^{\alpha_R}}(\mu_i)$, so by Lemma \ref{lem:b}
$s_i \in BR_{H}(\mu_i)$. Since $s_i$ was arbitrary, this proves that $R^{\alpha_R}$ is a fixpoint of $\overline{\GR}$.
Hence by Tarski's Fixpoint Theorem $R^{\alpha_R} \sse \overline{\textit{\GR}}^{\alpha_{\overline{\textit{\GR}}}}$.

Moreover, by Lemma \ref{lem:inc} $\overline{\textit{\GR}}^{\alpha_{\overline{\textit{\GR}}}} \sse R^{\alpha_R}$.
Consequently $R^{\alpha_R} = \overline{\textit{\GR}}^{\alpha_{\overline{\textit{\GR}}}}$.
\II

\NI
$(iii)$
We proceed  by transfinite induction.
Suppose the claim holds for all $\beta \leq \alpha$.
To prove it for $\alpha+1$ we first we prove that 
$\textit{\LR}^{\alpha + 1} \sse \textit{\LR}^{\alpha}$.
Let $s_i$ be a strategy of player $i$ in $\textit{\LR}^{\alpha+1}$.
Then for some
$\mu_{i} \in {\cal B}_{i} \stackrel{.}{\cap}  \textit{\LR}^{\alpha}$
we have $s_i \in BR_{\textit{\LR}^{\alpha}}(\mu_i)$.
By the induction hypothesis $\textit{\LR}^{\alpha} = \overline{\textit{\LR}}^{\alpha}$,
so by Lemma \ref{lem:b} $s_i \in BR_{H}(\mu_i)$.

Further, since $\overline{\textit{\LR}}$ is contracting,  
by property \textbf{A} for all $\beta \leq
\alpha$ we have $\mu_{i} \in {\cal B}_{i} \stackrel{.}{\cap}
\overline{\LR}^{\beta}$ and hence by the induction hypothesis
for all $\beta \leq
\alpha$ we have $\mu_{i} \in {\cal B}_{i} \stackrel{.}{\cap}
\textit{\LR}^{\beta}$.
So $s_i$ is a strategy of player $i$ in all
restrictions $\textit{\LR}^{\beta}$ for $\beta \leq \alpha$,
in particular $\textit{\LR}^{\alpha}$.
Since $s_i$ was arbitrary, this proves that 
$\textit{\LR}^{\alpha+1} \sse \textit{\LR}^{\alpha}$.

Hence
\[
\textit{\LR}^{\alpha +1} = \textit{\LR}^{\alpha +1} \cap\textit{\LR}^{\alpha} = \overline{\textit{\LR}}(\textit{\LR}^{\alpha}) = \overline{\textit{\LR}}^{\alpha +1},
\]
where the last equality holds by the induction hypothesis.

The induction step for a limit ordinal is immediate.
\HB
\VV

We mentioned already that property \textbf{B} is satisfied by the
compact and continuous games and pure belief structures.  So part $(i)$
is closely related to the result of \cite{Amb94} who showed that for
such games the finite iterations of his versions of
$\overline{\textit{\GR}}$ and $\overline{\textit{\LR}}$ operators
coincide for the limited case of two-player games and pure belief
structures. In his definition both operators are defined by
considering the reduction for each player in succession and not in
parallel.

In summary, by virtue of the results established in the last three sections,
in the presence of property \textbf{B}, rationalizability can be equivalently defined using any of the
introduced four operators $\textit{\GR}, \overline{\GR}, \textit{\LR}$  and $\overline{\LR}$.
However, when property \textbf{B} does not hold, only the first two operators are of interest.

\section{Strict dominance: the $\emph{\GS}$ and $\overline{\emph{\GS}}$ operators}
\label{sec:sd}

Recall that one of the consequences of the assumption that all players
are rational is that none of them could possibly use any strictly
dominated strategy. In this and the next section we apply our general
approach to operators on games to establish analogous results for the
notion strict dominance.\footnote{As is well-known there is an
  intimate connection between the notions of rationalizability and
  strict dominance. This, however, has no bearing on our results. We
  shall discuss this matter in Section \ref{sec:mixed}.}  In what
follows we analyze four operators that can be naturally used to define
iterated elimination of strictly dominated strategies.

Given a restriction $G := (S_1, \LL, S_n)$ of $H = (T_1, \LL, T_n, p_1, \LL, p_n)$
and two strategies $s_i, s'_i$ from $T_i$ we write
$s'_i \succ_{G} s_i$ as an abbreviation for
the statement $\fa s_{-i} \in S_{-i} \: p_{i}(s'_i, s_{-i}) > p_{i}(s_i, s_{-i})$
and say then that $s'_i$ \oldbfe{strictly dominates} $s_i$ \oldbfe{on} $G$
or, equivalently, that 
$s_i$ \oldbfe{is strictly dominated} \oldbfe{on} $G$ \oldbfe{by} $s'_i$.

First, we introduce the following operator $\textit{\GS}$ (standing for `global strict dominance')
on the set of restrictions of $H$:

\[
\textit{\GS}(G) := (S'_1, \LL, S'_n),
\]
where for all $i \in [1..n]$

\[
S'_i := \{ s_i \in T_i \mid \neg \te s'_i \in T_i \: s'_i \succ_{G} s_i\}.
\]

So $\textit{\GS}(G)$ is obtained by removing from $H$ all strategies
that are strictly dominated on $G$ by some strategy in $H$ and
\emph{not} in $G$.  The reasoning embodied in the definition of this operator
is analogous to the one used for $\textit{\GR}$: we remove a strategy from $H$
if a `better' (here: strictly dominating) strategy exists in the
initial game $H$, \emph{even} if in the iteration process leading
from $H$ to $G$ this better strategy might have been removed.

The \textit{\GS} operator was introduced in \cite{MR90}, where only its
iterations up to $\omega$ were considered.  As noted there this
operator is clearly monotonic. So Note \ref{not:contracting} applies
and we can confine our considerations to the $\overline{\textit{\GS}}$
operator (defined by $\overline{\textit{\GS}}(G) := \textit{\GS}(G) \cap
G$) that is also monotonic.\footnote{The reader may now notice that in
  Example \ref{exa:strict} we used the $\overline{\GS}$ operator
  instead of $\GS$, which is more complicated to justify informally.}
On the account of the results of Section
\ref{sec:reductions} this operator enjoys the same properties as the
$\overline{\GR}$ operator. We summarize them in the theorem below.

\begin{theorem}[$\overline{\textit{\GS}}$] \label{thm:sd}
\mbox{} \hspace{-8mm}

\begin{enumerate} \smallromani

\item The largest fixpoint of $\overline{\textit{\GS}}$ exists and is its outcome.

\item $\overline{\textit{\GS}}$ is order independent.
  
\item For all relaxations $R$ of $\overline{\textit{\GS}}$ and all $\alpha$ we have
  $\overline{\textit{\GS}}^{\alpha} \sse R^{\alpha}$.
\HB
\end{enumerate}
\end{theorem}

The $\overline{\textit{\GS}}$ operator was introduced and analyzed in
\cite{CLL05}, where the need for transfinite iterations was explained and
where properties $(i)$ and $(ii)$ were proved.  

\section{Strict dominance: the $\emph{\LS}$ and $\overline{\emph{\LS}}$ operators}
\label{sec:sd1}

The customary definition of the iterated elimination of strictly
dominated strategies (IESDS) involves elimination of strategies that
are strictly dominated by a strategy from the \emph{currently
  considered} game and \emph{not} the initial game $H$.  A natural
operator that formalizes this idea is the following one ($\textit{\LS}$
stands for `local strict dominance'):

\[
\textit{\LS}(G) := (S'_1, \LL, S'_n),
\]
where $G := (S_1, \LL, S_n)$ and for all $i \in [1..n]$

\[
S'_i := \{ s_i \in T_i \mid \neg \te s'_i \in S_i \: s'_i \succ_{G} s_i\}.
\]

So here, in contrast to $\textit{\GS}$, a strategy is removed if it is
strictly dominated on $G$ by some strategy \emph{in} $G$ itself.
However, the $\textit{\LS}$ operator is not acceptable for the same
reason as the previously considered $\textit{\LR}$ operator.

\begin{example}  \label{exa:1a}
  Reconsider the two-player game $H$ from Example \ref{exa:1} in which
  the set of strategies for each player is the set of natural numbers.
  The payoff to each player is the number he selected.  In this game
  each strategy is strictly dominated.  

So $\textit{\LS}(H) = (\ES,
  \ES)$. But $\textit{\LS}((\ES, \ES)) = H$, since the condition $\neg
  \te s'_i \in S_i \: s'_i \succ_{G} s_i$ is vacuously satisfied when
  $S_i = \ES$.  
So the iterations of $\textit{\LS}$ oscillate between
  $H$ and $(\ES, \ES)$.  Hence the outcome of $\textit{\LS}$ does not
  exist.

A fortiori $\textit{\LS}$ is not monotonic and
not contracting. The most problematic is of course that this operator yields a non-empty restriction (in fact, the initial
game) when applied to an empty restriction.
\HB
\end{example}

So, as in the case of the $\textit{\LR}$ operator, we confine our
attention to the contracting version $\overline{\LS}$ (defined by
$\overline{\LS}(G) := \textit{\LS}(G) \cap G$).  In fact, in the literature
the $\textit{\LS}$ operator was not considered but rather 
$\overline{\textit{\LS}}$.  However,
$\overline{\textit{\LS}}$ fails to be monotonic, even for finite games.
Indeed, the game from Example \ref{exa:012} provides the
evidence.

The $\overline{\textit{\LS}}$ operator was shown to be order
independent for finite games in \cite{GKZ90}.  It was studied for
infinite games in \cite{DS02} where it was noted that it fails to be
order independent for arbitrary games.  This is immediate to 
see using the game from Examples \ref{exa:1} and \ref{exa:1a}.
Indeed, the outcome of $\overline{\textit{\LS}}$ is then $(\ES, \ES)$, whereas
for any $i \geq  0$ the restriction $(\{i\}, \{i\})$ is the outcome of a relaxation
 $R$ of $\overline{\textit{\LS}}$ such that $R(H) = (\{i\}, \{i\})$.
 
 In the remainder of this section we prove a limited form of order
 independence of $\overline{\textit{\LS}}$. Then, in the next section,
 we compare our results with those of \cite{DS02}.  First, following
 \cite{DS02}, we consider the following property of the initial game $H$:

\begin{description}
\item[P]  For all relaxations $R$ of $\overline{\textit{\LS}}$ and all $k \geq
  0$, every strictly dominated strategy in $R^{k}$ has an undominated
  (`best') dominator.
\end{description}

This property limited to the initial game (so for $k = 0$)
was considered in \cite{MR96}.
We generalize this property to all ordinals and formalize it as follows:

\begin{description}

\item[C$(\alpha)$]
For all  relaxations $R$ of $\overline{\textit{\LS}}$ and all $i \in [1..n]$ 
\begin{tabbing}
$\fa s_i \in T_i \: ( $\=$\te s'_i \in T_i \: s'_i \succ_{R^{\alpha}} s_i \myra$ \\
\> $\te s^{*}_i \in T_i  \: (s^{*}_i \succ_{R^{\alpha}} s_i \A \neg \te s'_i \in T_i \: s'_i \succ_{R^{\alpha}} s^{*}_i)).$
\end{tabbing}

\end{description}

Note that all the quantifiers range over $T_i$ and not $S_i$.  So we
refer to strict dominance by a strategy in the initial game $H$ and
\emph{not} in the currently considered reduction $R^{\alpha}$.  In
what follows we shall rather use the following simpler property:

\begin{description}

\item[D$(\alpha)$]
For all  relaxations $R$ of $\overline{\textit{\LS}}$ and all $i \in [1..n]$

\[
\fa s_i \in T_{i}(
(\te s'_i \in T_i \: s'_i \succ_{R^{\alpha}} s_i) \myra 
\te s^{*}_i \in S_i \: s^{*}_i \succ_{R^{\alpha}} s_i),
\]
where $R^{\alpha} := (S_1, \LL, S_n)$.
\end{description}

It states that each $s_i$ strictly dominated in $R^{\alpha}$ 
is in fact strictly dominated in $R^{\alpha}$ by some strategy in $R^{\alpha}$.

First we establish a lemma that clarifies the relation between these two properties.

\begin{lemma} \label{lem:d}
For each $\alpha$, \textbf{C}($\alpha$) implies \textbf{D}($\alpha$).
\end{lemma}

\Proof
Consider a  relaxation $R$ of $\overline{\textit{\LS}}$, some $i \in [1..n]$. Let $R^{\alpha} := (S_1, \LL, S_n)$.
Suppose $s_i \in T_i$ and $s'_i \in T_i$ are such that
$s'_i \succ_{R^{\alpha}} s_i$.
By property \textbf{C}($\alpha$)
\[
\te s^{*}_i \in T_i \: (s^{*}_i \succ_{R^{\alpha}} s_i \A \neg \te s'_i \in T_i \: s'_i \succ_{R^{\alpha}} s^{*}_i).
\]
$\overline{\textit{\LS}}$ is contracting so for $\beta \leq \alpha$ we
have $R^{\alpha} \sse R^{\beta}$ and consequently $\fa \beta \leq
\alpha \: \neg \te s'_i \in T_i \: s'_i \succ_{R^{\beta}} s^{*}_i$.
Hence $s^{*}_i$ is a strategy of player $i$ in all restrictions
$R^{\beta}$ for $\beta \leq \alpha$.  In particular $s^{*}_i \in S_i$.
\HB
\VV

Consequently all the results that follow also hold when 
property \textbf{C}($\alpha)$ is used instead of
\textbf{D}($\alpha)$.

We now establish a number of consequences of property
\textbf{D}($\alpha$) for various values of $\alpha$.  Part $(i)$ is a
counterpart of Theorem \ref{thm:b}$(i)$.  Part $(ii)$ is a limited
order independence result for the
$\overline{\textit{\LS}}$ operator.  Finally, in part $(iii)$ we compare
the operators $\textit{\LS}$ and $\overline{\textit{\LS}}$ 
and establish an analogue of Theorem \ref{thm:b}$(iii)$.

\begin{theorem} \label{thm:equ2}
\mbox{} \hspace{-8mm}

\begin{enumerate} \smallromani

\item 
Assume property $\fa \beta < \alpha$\textbf{D}($\beta$)
for an ordinal $\alpha$.
Then $\overline{\textit{\GS}}^{\alpha} = \overline{\textit{\LS}}^{\alpha}$.
In particular,  if $\fa \alpha$\textbf{D}($\alpha$), then
$\overline{\textit{\GS}}^{\alpha_{\overline{\textit{\GS}}}} = \overline{\textit{\LS}}^{\alpha_{\overline{\textit{\LS}}}}$.

\item 
Assume property $\fa \alpha$\textbf{D}($\alpha$).
Then the $\overline{\textit{\LS}}$ operator is order independent.

\item 
Assume property $\fa \beta < \alpha + 1 \:$\textbf{D}($\beta$).
Then $\textit{\LS}^{\alpha} = \overline{\textit{\LS}}^{\alpha}$.
In particular, if $\fa \alpha$\textbf{D}($\alpha$), then
$\textit{\LS}^{\alpha_{\textit{\LS}}}$ exists and
$\textit{\LS}^{\alpha_{\textit{\LS}}} = 
\overline{\textit{\LS}}^{\alpha_{\overline{\textit{\LS}}}}$.

\end{enumerate}
\end{theorem}

\Proof

\NI
$(i)$ In view of Lemma \ref{lem:inc} we only need to prove the inclusion
$\overline{\textit{\LS}}^{\alpha} \sse \overline{\textit{\GS}}^{\alpha}$ and only the induction step
for a successor ordinal requires a justification.
So suppose that $\fa \beta < \alpha + 1 \:$\textbf{D}($\beta$). 
Let $\overline{\textit{\LS}}^{\alpha} := (S_1, \LL, S_n)$.

Let $s_i$ be a strategy of player $i$ in $\overline{\textit{\LS}}^{\alpha+1}$.
By definition $s_i \in S_i$ and 
$\neg \te s'_i \in S_i \: s'_i \succ_{\overline{\textit{\LS}}^{\alpha}} s_i$, so
by the assumed property
 $\neg \te s'_i \in T_i \: s'_i \succ_{\overline{\textit{\LS}}^{\alpha}} s_i$.
Hence $\neg \te s'_i \in T_i s'_i \succ_{\overline{\textit{\GS}}^{\alpha}} s_i$,
since by the induction hypothesis  $\overline{\textit{\LS}}^{\alpha} \sse \overline{\textit{\GS}}^{\alpha}$.
Also, because of the same inclusion, $s_i$ is a strategy of player $i$ in
$\overline{\textit{\GS}}^{\alpha}$. So $s_i$ is a strategy of player $i$ in $\overline{\textit{\GS}}^{\alpha+1}$.
\II

\NI
$(ii)$
The proof is similar to that of Theorem \ref{thm:b}$(ii)$.
Consider a relaxation $R$ of $\overline{\textit{\LS}}$
that has an outcome.
$R^{\alpha_R}$ is a fixpoint of $R$, so
$R^{\alpha_R}$ is also a fixpoint of $\overline{\textit{\LS}}$.

Suppose now $R^{\alpha_R} := (S_1, \LL, S_n)$ and take some
$s_i \in S_i$.
By the definition of $\overline{\textit{\LS}}$ we have
$\neg \te s^{*}_i \in S_i \: s^{*}_i \succ_{R^{\alpha_R}} s_i$, 
so by property \textbf{D}$(\alpha_R)$ we get
$\neg \te s'_i \in T_i \: s'_i \succ_{R^{\alpha_R}} s_i$.
Since $s_i$ was arbitrary, this proves that $R^{\alpha_R}$ is a fixpoint of $\overline{\textit{\GS}}$.
Hence by Tarski's Fixpoint Theorem $R^{\alpha_R} \sse \overline{\textit{\GS}}^{\alpha_{\overline{\textit{\GS}}}}$.

Moreover, for all $G$ we have both $\overline{\textit{\GS}}(G) \sse \overline{\textit{\LS}}(G)$ and
$\overline{\textit{\LS}}(G) \sse R(G)$, so by Lemma \ref{lem:inc}
$\overline{\textit{\GS}}^{\alpha_{\overline{\textit{\GS}}}} \sse R^{\alpha_R}$
and consequently $R^{\alpha_R} = \overline{\textit{\GS}}^{\alpha_{\overline{\textit{\GS}}}}$.
\II

\NI
$(iii)$
We prove the claim by transfinite induction.
Assume it holds for all $\beta \leq \alpha$. We prove it for $\alpha +1$.
So suppose that $\fa \beta < \alpha + 2 \:$\textbf{D}($\beta$). 
Let $\textit{\LS}^{\alpha} := (S_1, \LL, S_n)$.
Let $s_i$ be a strategy of player $i$ in $\textit{\LS}^{\alpha +1}$.
By definition 
\[
\neg \te s'_i \in S_i \: s'_i \succ_{\textit{\LS}^{\alpha}} s_i.
\]

By the induction hypothesis $\textit{\LS}^{\alpha} =
\overline{\LS}^{\alpha}$, so by the assumed property $\neg
\te s'_i \in T_i \: s'_i \succ_{\overline{\textit{\LS}}^{\alpha}}
s_i$.  So $s_i$ is a strategy of player $i$ in
$\textit{\GS}(\overline{\LS}^{\alpha})$.  But by part $(i)$
$\overline{\textit{\LS}}^{\alpha} = \overline{\textit{\GS}}^{\alpha}$
and $\overline{\textit{\LS}}^{\alpha+1} =
\overline{\textit{\GS}}^{\alpha+1}$.  Also, since $\textit{\GS}$ is
monotonic, by Note \ref{not:contracting}
$\overline{\textit{\GS}}^{\alpha} = \textit{\GS}^{\alpha}$ and
$\overline{\textit{\GS}}^{\alpha+1} = \textit{\GS}^{\alpha+1}$.  So
\[
\textit{\GS}(\overline{\textit{\LS}}^{\alpha}) = \textit{\GS}(\overline{\textit{\GS}}^{\alpha}) = 
\textit{\GS}^{\alpha+1} = \overline{\textit{\LS}}^{\alpha+1}.
\]

Since $s_i$ was arbitrary, this shows that 
$\textit{\LS}^{\alpha+1} \sse \overline{\textit{\LS}}^{\alpha+1}$.
But also

\[
\overline{\textit{\LS}}^{\alpha +1} = 
\overline{\textit{\LS}}(\textit{\LS}^{\alpha}) = 
\textit{\LS}^{\alpha +1} \cap\textit{\LS}^{\alpha} \sse
\textit{\LS}^{\alpha +1},
\]
where the first equality holds by the induction hypothesis.
So
$\overline{\textit{\LS}}^{\alpha +1} = \textit{\LS}^{\alpha +1}$.

The induction step for a limit ordinal is immediate.
\HB
\VV





Property $\fa \alpha$\textbf{C}($\alpha$) obviously holds when the initial
game is finite and implies by Lemma \ref{lem:d}
property $\fa \alpha$\textbf{D}($\alpha$), so
part $(ii)$ generalizes the already
mentioned result of \cite{GKZ90}.  Further, note that in the proof of
part $(i)$ we actually use a weaker property than $\fa \beta <
\alpha$\textbf{D}($\beta$). Indeed, we use it only
with $R$ equal to $\overline{\textit{\LS}}$, so it suffices to assume
that for all $\beta < \alpha$
\[
\fa s_i \in T_{i}(
(\te s'_i \in T_i \: s'_i \succ_{\overline{\textit{\LS}}^{\beta}} s_i) \myra 
\te s^{*}_i \in S_i \: s^{*}_i \succ_{\overline{\textit{\LS}}^{\beta}} s_i),
\]
where $\overline{\textit{\LS}}^{\beta} := (S_1, \LL, S_n)$.

In summary, the iterations of two operators, $\textit{\GS}$ and
$\overline{\textit{\GS}}$, always coincide and each of them is order
independent.  
Further, when $\fa
\alpha$\textbf{D}($\alpha$) holds, all iterations of all four operators, $\textit{\GS},
\overline{\textit{\GS}}, \textit{\LS}$ and $\overline{\textit{\LS}}$,
coincide and any of them can be used to define the outcome of IESDS.
The property $\fa \alpha$\textbf{D}($\alpha$) holds when the initial
game is finite, so to conclude this section let us summarize the consequences
of the above result for this case.

\begin{corollary} \label{cor:finite}
  Suppose the initial game $H$ is finite. Then 
  \begin{enumerate}\smallromani
  \item for all $k \geq 0$ the iterations of all four operators
    coincide, i.e., 
\[
\textit{\GS}^{k} =
    \overline{\textit{\GS}}^{k} = \overline{\textit{\LS}}^{k} = \textit{\LS}^{k},
\]

  \item the outcomes of these four operators exist and coincide,

\item the operators $\textit{\GS}$, $\overline{\textit{\GS}}$ and
  $\overline{\textit{\LS}}$ are order independent.  
\HB
  \end{enumerate}
\end{corollary}

\section{Comparison with the results of Dufwenberg and Stegeman}
\label{sec:comp}

We now compare the 
results of the previous section with those 
of \cite{DS02}. First, we introduce the following notion.  We say that
$G$ is an \oldbfe{$\omega$-outcome} of an operator $T$ on the set of
restrictions of $H$ if $G$ is an outcome of $T$ and $\alpha_T \leq
\omega$.\footnote{In \cite{DS02} an $\omega$-outcome is called a
  \oldbfe{maximal} ($\myra^{\hspace{-1mm} *}$)-\oldbfe{reduction}.}

That is, $G$ is an $\omega$-outcome of $T$ if $G = T^{\omega}$ and
$T^{\omega+1} = T^{\omega}$.  For a contracting operator, in contrast
to the outcome, an $\omega$-outcome does not need to exist.
The study of \cite{DS02} focuses on the set

\[
\mbox{$\omega(\overline{\textit{\LS}})$ := \{$G \mid G$ is an $\omega$-outcome of a relaxation of 
$\overline{\textit{\LS}}$\}}.
\]
If this set has at most one element, then they view $\overline{\textit{\LS}}$ as order independent. In what follows we refer then to
\oldbfe{DS-order independence}.
Recall that according to our definition an operator $T$ is order independent if the set 
\[
\mbox{\{$G \mid G$ is an outcome of a relaxation of $T$\}}
\]
has at most one element.
We can then state the main results of \cite{DS02} as follows.
To clarify part $(i)$ recall that by definition 
a restriction $(S_1, \LL, S_n)$ is a game if each $S_i$ is non-empty.

\begin{theorem} \label{thm:ds02}
\mbox{} \hspace{-8mm}

\begin{enumerate} \smallromani
  
\item If $H$ is compact and own-uppersemicontinuous and the set
  $\omega(\overline{\textit{\LS}})$ has an element which is a game, then this
  is its only element.

\item If $H$ is compact and continuous, 
then the set $\omega(\overline{\textit{\LS}})$ has precisely one element
and this element is a compact and continuous game.
\HB
\end{enumerate}
\end{theorem}

\cite{DS02} considered property \textbf{P} from the previous section
but formalized it as $\fa k < \omega$\textbf{E}$(k)$, where for an
ordinal $\alpha$ we have:

\begin{description}

\item[E$(\alpha)$]
For all  relaxations $R$ of $\overline{\textit{\LS}}$ and all $i \in [1..n]$ 
\begin{tabbing}
$\fa s_i \in S_i \: ( $\=$\te s'_i \in S_i \: s'_i \succ_{R^{\alpha}} s_i \myra$ \\
\> $\te s^{*}_i \in S_i  \: (s^{*}_i \succ_{R^{\alpha}} s_i \A \neg \te s'_i \in S_i \: s'_i \succ_{R^{\alpha}} s^{*}_i)),$
\end{tabbing}
where $R^{\alpha} := (S_1, \LL, S_n)$.
\end{description}

They showed that property $\fa k < \omega$\textbf{E}($k$) is satisfied
by the compact and own-uppersemicontinuous games but is not a
sufficient condition for DS-order independence of
$\overline{\textit{\LS}}$.  Note that the difference
between the properties \textbf{C}$(\alpha)$ and \textbf{E}$(\alpha)$
is that in the former all the quantifiers range over $T_i$ and not
$S_i$.  So in \textbf{C}$(\alpha)$ we refer to strict dominance by a
strategy in the initial game $H$, while in \textbf{E}$(\alpha)$ to 
strict dominance by a strategy in the currently considered reduction
$R^{\alpha}$.

The following result then relates property \textbf{D}$(\alpha)$
(and hence by Lemma \ref{lem:d} property \textbf{C}($\alpha$))
to DS-order independence of $\overline{\textit{\LS}}$.  

\begin{theorem}
  Assume property \textbf{D}($\omega$).  Then
  $\omega(\overline{\textit{\LS}}) \sse
  \{\overline{\textit{\LS}}^{\omega}\}$, so
$\overline{\textit{\LS}}$ is DS-order independent.
\end{theorem}
\Proof
Let $R$ be a relaxation of $\overline{\textit{\LS}}$ such that $\alpha_{R} \leq \omega$.
For all $\alpha$ we have both $\overline{\textit{\GS}}(R^{\alpha}) \sse \overline{\textit{\LS}}(R^{\alpha})$ and
$\overline{\textit{\LS}}(R^{\alpha}) \sse R(R^{\alpha})$, so by Lemma \ref{lem:inc}
\begin{equation}
  \label{equ:ssr}
\overline{\textit{\GS}}^{\omega} \sse R^{\omega}.  
\end{equation}

Let $R^{\omega} := (S_1, \LL, S_n)$.
Take a strategy $s_i \in S_i$.
$R^{\omega}$ is a fixpoint of $R$, so it
is a fixpoint of $\overline{\textit{\LS}}$. Hence
$\neg \te s'_i \in S_i \: s'_i \succ_{R^{\omega}} s_i$, so
by property \textbf{D}($\omega$)
$\neg \te s'_i \in T_i \: s'_i \succ_{R^{\omega}} s_i$.
Since $s_i$ was arbitrary, this shows that $R^{\omega}$ is a fixpoint of
$\overline{\textit{\GS}}$.
So by the $\overline{\textit{\GS}}$ Theorem \ref{thm:sd}$(i)$ we have
$R^{\omega} \sse \overline{\textit{\GS}}^{\alpha_{\overline{\textit{\GS}}}}$.
Also, $\overline{\textit{\GS}}$ is contracting, so 
$\overline{\textit{\GS}}^{\alpha_{\overline{\textit{\GS}}}} \sse
\overline{\textit{\GS}}^{\omega}$ and hence
\[
R^{\omega} \sse \overline{\textit{\GS}}^{\omega}.
\]
This inclusion combined with (\ref{equ:ssr}) yields 
$R^{\omega} = \overline{\textit{\GS}}^{\omega}$.
By Theorem \ref{thm:equ2}$(i)$ $R^{\omega} = \overline{\textit{\LS}}^{\omega}$.

So we showed that any $\omega$-outcome of a relaxation of 
$\overline{\textit{\LS}}$ equals $\overline{\textit{\LS}}^{\omega}$.
This concludes the proof.
\HB
\VV

\cite{DS02} also showed that for compact and own-upper\-semicontinuous games
property $\fa k \leq \omega$\textbf{D}($k$) holds.
So the above theorem is a minor improvement of the result of
\cite{DS02} listed earlier as Theorem \ref{thm:ds02}$(i)$. 
Finally, note that when $\fa k \leq \omega$\textbf{D}($k$) holds, then by
Theorem \ref{thm:equ2}$(i)$ and $(iii)$ the first $\omega$ iterations
of all four operators, $\textit{\GS}, \overline{\textit{\GS}},
\textit{\LS}$ and $\overline{\textit{\LS}}$, coincide.  So in
particular this is the case for the compact and
own-uppersemicontinuous games.

\section{Strict dominance by a mixed strategy}
\label{sec:mixed}

It is well-known that the notions of strict dominance and
best responses, and consequently the notions of iterated elimination of strictly
dominated strategies and of rationalizability,
are closely related. In this context one usually
considers strict dominance by a mixed strategy.  In section we review
this relationship and clarify to what extent the results of the
previous three sections can be extended to mixed strategies.

In \cite{Mou84} the class of so-called \oldbfe{nice games} is
introduced for which the notions of a best response to a point belief
(i.e., a joint pure strategy of the opponents) 
and of not being weakly dominated by a pure strategy coincide.
These are games $(T_1, \LL, T_n, p_1, \LL, p_n)$ in which each
strategy set $T_i$ is a compact and convex subset of ${\cal R}$ and
each payoff function $p_i$ is continuous and \oldbfe{strictly
  quasiconcave} w.r.t.~$T_i$, where the latter means that for all
$\alpha \in (0,1)$ and $s_i, s'_i \in T_i$ with $s_i \neq s'_i$ and
all $s_{-i} \in T_{-i}$
\[
\mbox{$p_i(s_i, s_{-i}) \geq p_i(s'_i, s_{-i})$ implies $p_i(\alpha s_i + (1 - \alpha) s'_i, s_{-i}) > p_i(s'_i, s_{-i})$.}
\]

In \cite{Zim06} it is clarified that for nice games the notions of weak dominance
and strict dominance coincide. This yields for nice games equivalence between
the notions of a best response and of not being strictly dominated, both w.r.t.~pure strategies.
This result is then generalized by assuming instead of strict quasiconcativity that
each payoff function $p_i$
is \oldbfe{quasiconcave} w.r.t.~$T_i$, which means that
for all $\alpha \in (0,1)$ and $s_i, s'_i \in T_i$ with $s_i \neq s'_i$ and all $s_{-i} \in T_{-i}$
\[
\mbox{$p_i(s_i, s_{-i}) > p_i(s'_i, s_{-i})$ implies $p_i(\alpha s_i +
  (1 - \alpha) s'_i, s_{-i}) > p_i(s'_i, s_{-i})$}
\]
and
\[
\mbox{$p_i(s_i, s_{-i}) = p_i(s'_i, s_{-i})$ implies $p_i(\alpha s_i + (1 - \alpha) s'_i, s_{-i}) \geq p_i(s'_i, s_{-i})$.}
\]
It is also shown that the equivalence does not need to hold anymore if the strategy sets $T_i$ are
subsets of ${\cal R}^2$ instead of ${\cal R}$.

In the case of finite games the notion of strict dominance extends in
the obvious to the case of mixed strategies.  By the result of
\cite{Pea84} in two-player finite games the notions of a best response
to a mixed strategy of the opponent and of not being strictly
dominated by a mixed strategy coincide. In \cite{OR94} this result is
presented as a result for $n$-players finite games where the best
response is defined w.r.t.~correlated mixed strategies.  More
recently, in \cite{Zim05}, both results were generalized to compact
games with bounded and continuous payoff functions.

In an arbitrary game $(T_1, \LL, T_n, p_1, \LL, p_n)$ the set of mixed
strategies $\Delta T_i$ of player $i$ is defined as the set of
\emph{all} probability measures on a given $\sigma$-algebra of subsets
of $T_i$.  In the case of compact games with continuous payoff
functions it is customary (as in \cite{Zim05}) to take the
$\sigma$-algebra of Borel sets.  The payoff functions are then
extended from pure to mixed strategies in a standard way using
integration.  To ensure that the payoffs remain finite the original
payoff functions are assumed to be bounded.

If the payoff functions are unbounded, an
alternative (used in Example \ref{exa:1}) is to define mixed
strategies as the probability measures with a finite support.
More general approaches for two-player games are studied in \cite{Tij75}.

In what follows we just assume that given the initial game $(T_1, \LL, T_n$, $p_1,
\LL, p_n)$ for each $i \in [1..n]$ a set $\Delta T_i \supseteq T_i$ of mixed strategies
of player $i$ is given and that each payoff function $p_i$ is
extended to $p_{i}: \Delta T_1 \times \LL \times \Delta T_n \myra {\cal R}$.
If the initial game is finite, we take the usual set of mixed strategies
and the customary extension of each payoff function.

Then the results of Section \ref{sec:sd1} can be directly adapted to the case of
strict dominance by a mixed strategy as follows.
First, we introduce the counterpart of the 
$\LS$ operator defined by:

\[
\MLS(G) := (S'_1, \LL, S'_n),
\]
where $G := (S_1, \LL, S_n)$ and for all $i \in [1..n]$

\[
S'_i := \{ s_i \in T_i \mid \neg \te m_i \in \Delta S_i \: m_i \succ_{G} s_i\},
\]
where we use the extension of the $\succ_{G}$ relation to the pairs of mixed and pure strategies.

In the literature, in the case of finite games, the iterated
elimination of strategies that are strictly dominated by a mixed
strategy is defined as the iteration of $\overline{\MLS}$, the
contracting version of the above operator.  The obvious modification
of the \textit{\GS} operator to the mixed strategies is defined by:

\[
\MGS(G) := (S'_1, \LL, S'_n),
\]
where for all $i \in [1..n]$
\[
S'_i := \{ s_i \in T_i \mid \neg \te m_i \in \Delta T_i \: m_i \succ_{G} s_i\}.
\]

Just as $\GS$, the $\MGS$ operator is clearly monotonic.
Its contracting version,
$\overline{\MGS}$, is studied in
\cite{BFK06} (it is their map $\Phi$), where its relation to the
concept of best response sets is clarified.

Next, we modify the property \textbf{D}($\alpha$) to the case of mixed strategies:

\begin{description}

\item[MD$(\alpha)$]
For all relaxations $R$ of $\overline{\textit{\MLS}}$ and all $i \in [1..n]$
\[
\fa s_i \in T_{i}(
(\te m_i \in \Delta T_i \: m_i \succ_{R^{\alpha}} s_i) \myra 
\te m^{*}_i \in \Delta S_i \: m^{*}_i \succ_{R^{\alpha}} s_i),
\]
where $R^{\alpha} := (S_1, \LL, S_n)$.
\end{description}

Then we have the following direct counterpart of Theorem \ref{thm:equ2}.

\begin{theorem} \label{thm:equ3}
\mbox{} \hspace{-8mm}

\begin{enumerate} \smallromani

\item 
Assume property $\fa \beta < \alpha$\textbf{MD}($\beta$)
for an ordinal $\alpha$.
Then $\overline{\textit{\MGS}}^{\alpha} = \overline{\textit{\MLS}}^{\alpha}$.
In particular,  if $\fa \alpha$\textbf{MD}($\alpha$), then
$\overline{\textit{\MGS}}^{\alpha_{\overline{\textit{\MGS}}}} = \overline{\textit{\MLS}}^{\alpha_{\overline{\textit{\MLS}}}}$.

\item 
Assume property $\fa \alpha$\textbf{MD}($\alpha$).
Then the $\overline{\textit{\MLS}}$ operator is order independent.

\item 
Assume property $\fa \beta < \alpha + 1 \:$\textbf{MD}($\beta$).
Then $\textit{\MLS}^{\alpha} = \overline{\textit{\MLS}}^{\alpha}$.
In particular, if $\fa \alpha$\textbf{MD}($\alpha$), then
$\textit{\MLS}^{\alpha_{\textit{\MLS}}}$ exists and
$\textit{\MLS}^{\alpha_{\textit{\MLS}}} = 
\overline{\textit{\MLS}}^{\alpha_{\overline{\textit{\MLS}}}}$.

\end{enumerate}
\end{theorem}

\Proof
Analogous to the proof of Theorem \ref{thm:equ2} and omitted.
\HB
\VV

Consequently, when $\fa \alpha$\textbf{MD}($\alpha$) all iterations of
all four operators, $\textit{\MGS}$, $\overline{\textit{\MGS}},
\textit{\MLS}$ and $\overline{\textit{\MLS}}$, coincide. To establish
this result, as in the case of strict dominance by a pure strategy, it
is sufficient to use property \textbf{MD}($\alpha$) with $R$ equal to
$\overline{\textit{\MLS}}$.

The question remains for which games property \textbf{MD}($\alpha$)
holds.  We found (the details are relegated to another paper) that in
the case of finite games and the customary set of mixed strategies,
property \textbf{MD}($k$) holds for all $k \geq 0$. Since the closure
ordinals of the above four operators are then finite, we conclude
by the above theorem that all iterations of the above four operators then coincide.

\section{Concluding remarks}
\label{sec:conc}

In this paper we analyzed two widely used ways of reducing strategic
games concerned with the concepts of rationalizability and iterated
elimination of strictly dominated strategies. We showed that both
concepts can be defined in a number of ways that differ for arbitrary
infinite games.  Also, we clarified for which games these differences
disappear.  Our analysis was based on a general study of operators on
complete lattices and showed that concepts defined by means of monotonic
operators are easier to assess and study.

In some circumstances a reduction notion defined using a non-monotonic
operator still can be analyzed using our approach, by relating the
operator to a monotonic one. This is for example how we established the order
independence of the $\overline{\textit{\LS}}$ operator under the
assumption $\fa \alpha$\textbf{D}($\alpha$) ---by relating it to the
$\overline{\textit{\GS}}$ operator which is monotonic.

An important aspect of our analysis is that we allow transfinite
iterations of the corresponding operators.  Their use in an analysis
of reasoning used by rational agents can be baffling. These matters
were originally discussed in \cite{Lip91}, where a need for
transfinite iterations in the definition of rationalizability was
noted.  The distinction between finitary and infinitistic forms of
reasoning is well understood in mathematical logic. Both forms have
been widely used and we see no reason for limiting the study of games
to finitary methods.

The final matter that merits attention is the striking difference
between the way rationalizability and iterated elimination of strictly
dominated strategies (IESDS) have been traditionally defined and used
in the literature.  Both concepts are supposed to capture reasoning
used by rational players.  Yet, in the definition of the former
concept, according to \cite{Ber84}, the reference point for a deletion
of a strategy is the \emph{initial} game, while in the definition of
the latter one the \emph{currently considered} game.  This difference
is important, since the first approach yields a monotonic operator,
while the second one not.  


\section*{Acknowledgements}

We thank Xiao Luo for an interesting discussion on the subject of
\cite{Apt05} and \cite{CLL05}, references to \cite{MR90} and
\cite{Rit02}, and helpful suggestions.  Adam Brandenburger helped us
in assessing and framing the relevant results.  Also, we thank the
referees of earlier versions of this paper for useful comments.  In
particular, one of the referees of \cite{Apt05} drew our attention to
the paper \cite{Amb94}. Another referee spotted an error in one of our
examples.  Finally, one of the
referees of this submission provided most helpful, extensive, comments
and suggested Section \ref{sec:mixed}.

\bibliography{/ufs/apt/bib/e,/ufs/apt/bib/apt}
\bibliographystyle{handbk}

\end{document}

%% file: ecmds.tex
\newcommand{\oldbfe}[1]{\begin{bfseries}\emph{#1}\end{bfseries}}

\newcommand{\ES}{\mbox{$\emptyset$}}

\newcommand{\myra}{\mbox{$\:\rightarrow\:$}}

\newcommand{\A}{\mbox{$\ \wedge\ $}}

\newcommand{\sse}{\mbox{$\:\subseteq\:$}}

\newcommand{\fa}{\mbox{$\forall$}}
\newcommand{\te}{\mbox{$\exists$}}

\newcommand{\LL}{\mbox{$\ldots$}}

\newcommand{\C}[1]{\mbox{$\{{#1}\}$}}           

\newcommand{\NI}{\noindent}
\newcommand{\HB}{\hfill{$\Box$}}
\newcommand{\VV}{\vspace{5 mm}}

\newcommand{\II}{\vspace{2 mm}}

\newcommand{\szkew}[1]{\relax \setbox0=\hbox{\kern -24pt $\displaystyle#1$\kern 0pt }%
\box0}
{\catcode`\@=11 \global\let\ifjusthvtest@=\iffalse}

\newcounter{oldmycaption}



%% file: minusone.pstex_t
\begin{picture}(0,0)%
\includegraphics{minusone.pstex}%
\end{picture}%
\setlength{\unitlength}{2763sp}%
\begingroup\makeatletter\ifx\SetFigFont\undefined%
\gdef\SetFigFont#1#2#3#4#5{%
  \reset@font\fontsize{#1}{#2pt}%
  \fontfamily{#3}\fontseries{#4}\fontshape{#5}%
  \selectfont}%
\fi\endgroup%
\begin{picture}(6789,2161)(3112,-3710)
\put(3451,-3361){\makebox(0,0)[rb]{\smash{{\SetFigFont{8}{9.6}{\familydefault}{\mddefault}{\updefault}{\color[rgb]{0,0,0}$0$}%
}}}}
\put(3451,-3061){\makebox(0,0)[rb]{\smash{{\SetFigFont{8}{9.6}{\familydefault}{\mddefault}{\updefault}{\color[rgb]{0,0,0}$1$}%
}}}}
\put(3451,-2761){\makebox(0,0)[rb]{\smash{{\SetFigFont{8}{9.6}{\familydefault}{\mddefault}{\updefault}{\color[rgb]{0,0,0}$2$}%
}}}}
\put(3451,-2461){\makebox(0,0)[rb]{\smash{{\SetFigFont{8}{9.6}{\familydefault}{\mddefault}{\updefault}{\color[rgb]{0,0,0}$3$}%
}}}}
\put(3451,-2161){\makebox(0,0)[rb]{\smash{{\SetFigFont{8}{9.6}{\familydefault}{\mddefault}{\updefault}{\color[rgb]{0,0,0}$4$}%
}}}}
\put(3451,-1861){\makebox(0,0)[rb]{\smash{{\SetFigFont{8}{9.6}{\familydefault}{\mddefault}{\updefault}{\color[rgb]{0,0,0}$\ldots$}%
}}}}
\put(6001,-1861){\makebox(0,0)[rb]{\smash{{\SetFigFont{8}{9.6}{\familydefault}{\mddefault}{\updefault}{\color[rgb]{0,0,0}$\ldots$}%
}}}}
\put(6001,-2161){\makebox(0,0)[rb]{\smash{{\SetFigFont{8}{9.6}{\familydefault}{\mddefault}{\updefault}{\color[rgb]{0,0,0}$4$}%
}}}}
\put(6001,-2461){\makebox(0,0)[rb]{\smash{{\SetFigFont{8}{9.6}{\familydefault}{\mddefault}{\updefault}{\color[rgb]{0,0,0}$3$}%
}}}}
\put(6001,-2761){\makebox(0,0)[rb]{\smash{{\SetFigFont{8}{9.6}{\familydefault}{\mddefault}{\updefault}{\color[rgb]{0,0,0}$2$}%
}}}}
\put(6001,-3061){\makebox(0,0)[rb]{\smash{{\SetFigFont{8}{9.6}{\familydefault}{\mddefault}{\updefault}{\color[rgb]{0,0,0}$1$}%
}}}}
\put(6001,-3361){\makebox(0,0)[rb]{\smash{{\SetFigFont{8}{9.6}{\familydefault}{\mddefault}{\updefault}{\color[rgb]{0,0,0}$0$}%
}}}}
\put(7351,-1861){\makebox(0,0)[rb]{\smash{{\SetFigFont{8}{9.6}{\familydefault}{\mddefault}{\updefault}{\color[rgb]{0,0,0}$\ldots$}%
}}}}
\put(7351,-2161){\makebox(0,0)[rb]{\smash{{\SetFigFont{8}{9.6}{\familydefault}{\mddefault}{\updefault}{\color[rgb]{0,0,0}$4$}%
}}}}
\put(7351,-2461){\makebox(0,0)[rb]{\smash{{\SetFigFont{8}{9.6}{\familydefault}{\mddefault}{\updefault}{\color[rgb]{0,0,0}$3$}%
}}}}
\put(7351,-2761){\makebox(0,0)[rb]{\smash{{\SetFigFont{8}{9.6}{\familydefault}{\mddefault}{\updefault}{\color[rgb]{0,0,0}$2$}%
}}}}
\put(7351,-3061){\makebox(0,0)[rb]{\smash{{\SetFigFont{8}{9.6}{\familydefault}{\mddefault}{\updefault}{\color[rgb]{0,0,0}$1$}%
}}}}
\put(7351,-3361){\makebox(0,0)[rb]{\smash{{\SetFigFont{8}{9.6}{\familydefault}{\mddefault}{\updefault}{\color[rgb]{0,0,0}$0$}%
}}}}
\put(9901,-1861){\makebox(0,0)[rb]{\smash{{\SetFigFont{8}{9.6}{\familydefault}{\mddefault}{\updefault}{\color[rgb]{0,0,0}$\ldots$}%
}}}}
\put(9901,-2161){\makebox(0,0)[rb]{\smash{{\SetFigFont{8}{9.6}{\familydefault}{\mddefault}{\updefault}{\color[rgb]{0,0,0}$4$}%
}}}}
\put(9901,-2461){\makebox(0,0)[rb]{\smash{{\SetFigFont{8}{9.6}{\familydefault}{\mddefault}{\updefault}{\color[rgb]{0,0,0}$3$}%
}}}}
\put(9901,-2761){\makebox(0,0)[rb]{\smash{{\SetFigFont{8}{9.6}{\familydefault}{\mddefault}{\updefault}{\color[rgb]{0,0,0}$2$}%
}}}}
\put(9901,-3061){\makebox(0,0)[rb]{\smash{{\SetFigFont{8}{9.6}{\familydefault}{\mddefault}{\updefault}{\color[rgb]{0,0,0}$1$}%
}}}}
\put(3451,-3661){\makebox(0,0)[rb]{\smash{{\SetFigFont{8}{9.6}{\familydefault}{\mddefault}{\updefault}{\color[rgb]{0,0,0}$-1$}%
}}}}
\put(6001,-3661){\makebox(0,0)[rb]{\smash{{\SetFigFont{8}{9.6}{\familydefault}{\mddefault}{\updefault}{\color[rgb]{0,0,0}$-1$}%
}}}}
\put(9901,-3361){\makebox(0,0)[rb]{\smash{{\SetFigFont{8}{9.6}{\familydefault}{\mddefault}{\updefault}{\color[rgb]{0,0,0}$0$}%
}}}}
\end{picture}%

%% file: ratio07.bbl
\begin{thebibliography}{1999}

\bibitem[Aczel:\nameindex{Aczel, P.}:1977]{Acz77}
{\sc P.~Aczel\nameindex{Aczel, P.}}, An introduction to inductive definitions,
  in: {\em Handbook of Mathematical Logic}, J.~Barwise\nameindex{Barwise, J.},
  ed., North-Holland, pp.~739--782.

\bibitem[Ambroszkiewicz:\nameindex{Ambroszkiewicz, S.}:1994]{Amb94}
{\sc S.~Ambroszkiewicz\nameindex{Ambroszkiewicz, S.}}, Knowledge and best
  responses in games, {\em Annals of Operations Research}, 51, pp.~63--71.

\bibitem[Apt:\nameindex{Apt, K.~R.}:2005]{Apt05}
{\sc K.~R. Apt\nameindex{Apt, K.~R.}}, Order independence and
  rationalizability, in: {\em Proc. 10th Conference on Theoretical Aspects of
  Reasoning about Knowledge (TARK)}, Singapore, pp.~22--38.
\newblock Available from \url{http://portal.acm.org}.

\bibitem[Aumann:\nameindex{Aumann, R.}:1985]{Aum85}
{\sc R.~Aumann\nameindex{Aumann, R.}}, What is game theory trying to
  accomplish?, in: {\em Frontiers of Economics}, K.~Arrow\nameindex{Arrow, K.}
  and S.~Honkapohja\nameindex{Honkapohja, S.}, eds., Basil Blackwell,
  pp.~28--76.

\bibitem[Bernheim:\nameindex{Bernheim, B.~D.}:1984]{Ber84}
{\sc B.~D. Bernheim\nameindex{Bernheim, B.~D.}}, Rationalizable strategic
  behavior, {\em Econometrica}, 52, pp.~1007--1028.

\bibitem[Binmore:\nameindex{Binmore, K.}:1991]{Bin91}
{\sc K.~Binmore\nameindex{Binmore, K.}}, {\em Fun and Games: A Text on Game
  Theory}, D.C. Heath.

\bibitem[Binmore:\nameindex{Binmore, K.}:2007]{Bin07}
{\sc K.~Binmore\nameindex{Binmore, K.}}, {\em Playing for Real: A Text on Game
  Theory}, Oxford University Press, Oxford.

\bibitem[Brandenburger and Dekel:\nameindex{Brandenburger, A.}\nameindex{Dekel,
  E.}:1987]{BD87}
{\sc A.~Brandenburger\nameindex{Brandenburger, A.} and
  E.~Dekel\nameindex{Dekel, E.}}, Rationalizability and correlated equilibria,
  {\em Econometrica}, 55, pp.~1391--1402.

\bibitem[Brandenburger, Friedenberg and Keisler:\nameindex{Brandenburger,
  A.}\nameindex{Friedenberg, A.}\nameindex{Keisler, H.}:2006]{BFK06}
{\sc A.~Brandenburger\nameindex{Brandenburger, A.},
  A.~Friedenberg\nameindex{Friedenberg, A.}, and H.~Keisler\nameindex{Keisler,
  H.}}, Fixed points for strong and weak dominance.
\newblock Working paper. Available from
  \verb?http://pages.stern.nyu.edu/~abranden?/.

\bibitem[Chen, Long and Luo:\nameindex{Chen, Y.-C.}\nameindex{Long,
  N.~V.}\nameindex{Luo, X.}:2005]{CLL05}
{\sc Y.-C. Chen\nameindex{Chen, Y.-C.}, N.~V. Long\nameindex{Long, N.~V.}, and
  X.~Luo\nameindex{Luo, X.}}, {\em Iterated strict dominance in general games}.
\newblock Available from \url{http://www.sinica.edu.tw/~xluo/pa10.pdf}.

\bibitem[Dufwenberg and Stegeman:\nameindex{Dufwenberg, M.}\nameindex{Stegeman,
  M.}:2002]{DS02}
{\sc M.~Dufwenberg\nameindex{Dufwenberg, M.} and
  M.~Stegeman\nameindex{Stegeman, M.}}, Existence and uniqueness of maximal
  reductions under iterated strict dominance, {\em Econometrica}, 70,
  pp.~2007--2023.

\bibitem[Gilboa, Kalai and Zemel:\nameindex{Gilboa, I.}\nameindex{Kalai,
  E.}\nameindex{Zemel, E.}:1990]{GKZ90}
{\sc I.~Gilboa\nameindex{Gilboa, I.}, E.~Kalai\nameindex{Kalai, E.}, and
  E.~Zemel\nameindex{Zemel, E.}}, On the order of eliminating dominated
  strategies, {\em Operation Research Letters}, 9, pp.~85--89.

\bibitem[Heifetz and Samet:\nameindex{Heifetz, A.}\nameindex{Samet,
  D.}:1998]{HS98}
{\sc A.~Heifetz\nameindex{Heifetz, A.} and D.~Samet\nameindex{Samet, D.}},
  Knowledge spaces with arbitrarily high rank, {\em Games and Economic
  Behavior}, 22, pp.~260--273.

\bibitem[Herings and Vannetelbosch:\nameindex{Herings,
  P.~J.-J.}\nameindex{Vannetelbosch, V.~J.}:2000]{HV00}
{\sc P.~J.-J. Herings\nameindex{Herings, P.~J.-J.} and V.~J.
  Vannetelbosch\nameindex{Vannetelbosch, V.~J.}}, The equivalence of the
  {Dekel-Fudenberg} iterative procedure and weakly perfect rationalizability,
  {\em Economic Theory}, pp.~677--687.

\bibitem[Knaster:\nameindex{Knaster, B.}:1928]{Kna28}
{\sc B.~Knaster\nameindex{Knaster, B.}}, Un th\'{e}or\`{e}me sur les fonctions
  d'ensembles, {\em Annales de la Societe Polonaise de Mathematique}, 6,
  pp.~133--134.

\bibitem[Lipman:\nameindex{Lipman, B.~L.}:1991]{Lip91}
{\sc B.~L. Lipman\nameindex{Lipman, B.~L.}}, How to decide how to decide how to
  $\dots$: Modeling limited rationality, {\em Econometrica}, 59,
  pp.~1105--1125.

\bibitem[Lipman:\nameindex{Lipman, B.~L.}:1994]{Lip94}
{\sc B.~L. Lipman\nameindex{Lipman, B.~L.}}, A note on the implications of
  common knowledge of rationality, {\em Games and Economic Behavior}, 6,
  pp.~114--129.

\bibitem[Milgrom and Roberts:\nameindex{Milgrom, P.}\nameindex{Roberts,
  J.}:1990]{MR90}
{\sc P.~Milgrom\nameindex{Milgrom, P.} and J.~Roberts\nameindex{Roberts, J.}},
  Rationalizability, learning, and equilibrium in games with strategic
  complementarities, {\em Econometrica}, 58, pp.~1255--1278.

\bibitem[Milgrom and Roberts:\nameindex{Milgrom, P.}\nameindex{Roberts,
  J.}:1996]{MR96}
{\sc P.~Milgrom\nameindex{Milgrom, P.} and J.~Roberts\nameindex{Roberts, J.}},
  Coalition-proofness and correlation with arbitrary communication
  possibilities, {\em Games and Economic Behavior}, 17, pp.~113--128.

\bibitem[Moulin:\nameindex{Moulin, H.}:1984]{Mou84}
{\sc H.~Moulin\nameindex{Moulin, H.}}, Dominance solvability and cournot
  stability, {\em Mathematical Social Sciences}, 7, pp.~83--102.

\bibitem[Osborne and Rubinstein:\nameindex{Osborne,
  M.~J.}\nameindex{Rubinstein, A.}:1994]{OR94}
{\sc M.~J. Osborne\nameindex{Osborne, M.~J.} and
  A.~Rubinstein\nameindex{Rubinstein, A.}}, {\em A Course in Game Theory}, The
  {MIT} Press, Cambridge, Massachusetts.

\bibitem[Pearce:\nameindex{Pearce, D.~G.}:1984]{Pea84}
{\sc D.~G. Pearce\nameindex{Pearce, D.~G.}}, Rationalizable strategic behavior
  and the problem of perfection, {\em Econometrica}, 52, pp.~1029--1050.

\bibitem[Ritzberger:\nameindex{Ritzberger, K.}:2001]{Rit02}
{\sc K.~Ritzberger\nameindex{Ritzberger, K.}}, {\em Foundations of
  Non-cooperative Game Theory}, Oxford University Press, Oxford.

\bibitem[Tarski:\nameindex{Tarski, A.}:1955]{Tar55}
{\sc A.~Tarski\nameindex{Tarski, A.}}, A lattice-theoretic fixpoint theorem and
  its applications, {\em Pacific J. Math}, 5, pp.~285--309.

\bibitem[Tijs:\nameindex{Tijs, S.}:1975]{Tij75}
{\sc S.~Tijs\nameindex{Tijs, S.}}, {\em Semi-infinite and infinite matrix games
  and bimatric games}, PhD thesis, Katholieke Universiteit Nijmegen.

\bibitem[Zimper:\nameindex{Zimper, A.}:2005]{Zim05}
{\sc A.~Zimper\nameindex{Zimper, A.}}, Equivalence between best responses and
  undominated strategies: a generalization from finite to compact strategy
  sets, {\em Economics Bulletin}, 3, pp.~1--6.

\bibitem[Zimper:\nameindex{Zimper, A.}:2006]{Zim06}
{\sc A.~Zimper\nameindex{Zimper, A.}}, A note on the equivalence of
  rationalizability concepts in generalized nice games, {\em International Game
  Theory Review}, 8, pp.~669--674.

\end{thebibliography}
